\documentclass[11pt]{article}
\usepackage{amssymb}
\usepackage[dvips]{graphicx}
\def\dd{\mbox{d}}

\def\o{\omega}
\def\bra{\langle}
\def\ket{\rangle}
\def\a{\alpha}
\def\b{\beta}
\def\d{\delta}
\def\D{\Delta}

\def\g{\gamma}
\def\G{\Gamma}
\def\e{\epsilon}
\def\ve{\varepsilon}

\def\f{\phi}
\def\F{\Phi}
\def\vf{\varphi}
\def\k{\kappa}
\def\l{\lambda}
\def\L{\Lambda}
\def\m{\mu}
\def\n{\nu}
\def\s{\sigma}
\def\S{\Sigma}
\def\o{\omega}

\def\r{\rho}
\def\t{\tau}
\def\th{\theta}

\def\pa{\partial}

\newcommand{\ti}[1]{\tilde{#1}}

\newcommand{\sm}[1]{\mbox{\scriptsize #1}}
\newcommand{\tn}[1]{\mbox{\tiny #1}}
\renewcommand{\@}[1]{\sqrt{#1}}
\newcommand{\Tr}{{\mbox{Tr}}\,}
\renewcommand{\le}[1]{\label{#1}\end{eqnarray}}
\newcommand{\be}{\begin{equation}}
\newcommand{\ee}{\end{equation}}
\newcommand{\bea}{\begin{eqnarray}}
\newcommand{\eea}{\end{eqnarray}}
\newcommand{\nn}{\nonumber}

\newcommand{\eq}[1]{(\ref{#1})}
\def\nn{\nonumber\\}

\def\ffract#1#2{\raise .35 em\hbox{$\scriptstyle#1$}\kern-.25em/
\kern-.2em\lower .22 em \hbox{$\scriptstyle#2$}}

\def\half{{1\over2}\,}
\begin{document}

\begin{flushright}
AEI-2004-123\\
{\tt hep-th/0412110}\\
\end{flushright}
\vskip0.1truecm

\begin{center}
\vskip 2.5truecm {\Large \textbf{Chern-Simons Theory, 2d Yang-Mills, and \\
$ $\\
Lie Algebra Wanderers}}
%Chern-Simons Theory and Brownian Motion on Lie Algebras}}
% as a Dissipative System}}
%Chern-Simons Theory, Brownian Motion, and the Stieltjes-Wigert Polynomials
%Chern-Simons as a Statistical Theory
\vskip 1truecm
%\vfill

{\large\bf{Sebastian de Haro}}\\
\vskip 1.5truecm {\it Max-Planck-Institut f\"{u}r
Gravitationsphysik\\
Albert-Einstein-Institut\\
14476 Golm, Germany}\\
\tt{sdh@aei.mpg.de}

\end{center}

\vskip 2truecm

\begin{center}
\textbf{\large Abstract}
\end{center}

We work out the relation between Chern-Simons, 2d Yang-Mills on the cylinder, and Brownian motion. We show that for the unitary, orthogonal and symplectic groups, various observables in Chern-Simons theory on $S^3$ and lens spaces are exactly given by counting the number of paths of a Brownian particle wandering in the fundamental Weyl chamber of the corresponding Lie algebra. We construct a fermionic formulation of Chern-Simons on $S^3$ which allows us to identify the Brownian particles as B-model branes moving on a non-commutative two-sphere, and construct 1- and 2-matrix models to compute Brownian motion ensemble averages.

\newpage

\tableofcontents

\newpage

\section{Introduction}

The idea that the large $N$ limit of quantum field theories gives closed string theories \cite{thooft,maldacena} has gained much in
tractability in the context of topological strings, where, as Gopakumar and Vafa
\cite{GoVa} showed --in the case of the conifold geometric transition-- the associated gauge 
theory is a {\it bosonic}, topological gauge theory, namely three-dimensional Chern-Simons theory on $S^3$. Recently, it has become more and 
more clear 
that other low-dimensional bosonic theories may also play a central role in string theory computations, and in particular, building on the old 
idea of Gross \cite{gross}, 2d Yang-Mills \cite{vafa,sdh,aosv}. Furthermore, these bosonic theories can often be mapped to statistical mechanical 
systems \cite{brownian,sdh,orv}, which provides one with useful computational tools \cite{matsuo}. Also, it is by now clear that Chern-Simons and 
two-dimensional Yang-Mills (in its usual and its quantum deformed versions) are closely connected 
\cite{sdh,aosv,br}, generalizing previous work 
which related them at zero coupling \cite{witten2d,blauthompson,polyakreshetikhin} (see also 
\cite{gorskynikita}).

In this paper we work out the relation between Chern-Simons, 2d Yang-Mills on the cylinder, and Brownian motion found in \cite{brownian,sdh}. We 
show that for unitary, orthogonal and symplectic groups, observables in Chern-Simons theory on $S^3$ are exactly given by counting the number of 
paths of a Brownian particle wandering in the fundamental Weyl chamber of the corresponding algebra, or equivalently by a number of non-intersecting movers on a line where certain boundary conditions are imposed. In particular, we compute the partition function, the expectation value of the unknot, and the expectation value of the Hopf link. Our results suggest that Brownian motion might give a rather general and independent way of computing knot and 3-manifold invariants.

We also comment on the relation between 2d Yang-Mills on the cylinder and Brownian motion in an affine Weyl chamber, already worked out in 
\cite{sdh}. This gave a connection between 2d Yang-Mills on the cylinder and Chern-Simons on lens spaces that made possible the computation of 
the modular transformation properties of 2d Yang-Mills on the cylinder, which for particular external states reduces to the quantum-deformed 2d 
Yang-Mills on the sphere. Thus, these can be translated into modular transformation properties of certain A-model amplitudes.

We construct a fermionic formulation of Chern-Simons theory which allows us to identify the Brownian particles with B-model branes moving on a non-commutative sphere.

Finally, we show how hermitian 1- and 2-matrix models can be used to compute Brownian motion observbles.

This is the companion paper of \cite{brownian}. Sections 2 and 3 include pedagogical introductions to Brownian motion and Chern-Simons theory, respectively. In sections 3, 4 and 5 we work out the relation between Chern-Simons and Brownian motion in various cases. In section 5 we discuss 2d Yang-Mills, in section 6 we work out the fermionic description of Chern-Simons and its relation to topological strings, and in section 7 we introduce matrix models for Brownian motion. In the appendices we give technical details, including a discussion of framing and a path integral counting of the Brownian motion paths.

\section{Random walks and Brownian motion}

Almost a hundred years ago, Einstein gave the first mathematical explanation of
the phenomenon of `Brownian motion', the zig-zag-like random motion 
characteristic of pollen grains suspended in water, observed by the botanist 
Robert Brown around 1827. Einstein's model was a discrete `random walk' model, 
that we will briefly review.

\subsection{Random walks}

In its simplest guise, we can regard the random walk of a particle as follows.
At every tick of the clock, the particle can jump either right or left with
equal probability. We want to compute the probabilty that after time $t$ the
particle will have traveled a distance $x$. We set $t=n\tau$, where $n$
denotes the total number of steps and $\t$ the average time between one step
and the next one; and $x=m l$, where $m$ is the number of steps in the 
positive $x$-direction minus the number of negative steps. The probability that
after $n$ steps the particle has made $m$ steps in the $x$-direction is then:
\be\label{discrete}
P(n|m)={n!\over[\half(n+m)]![\half(n-m)]!}\,{1\over 2^n}~,
\ee
where the binomial distribution comes from the fact that the particle takes a
total of $\half(n+m)$ steps in the positive $x$-direction, and $\half(n-m)$ in 
the negative direction. Of course we can rewrite this probability in terms of
$x$ and $t$ and take the limit of small $\t$, $l$. If we do this while 
keeping
\be
D=l^2/2\t
\ee
fixed, applying Stirling's formula we get for the probability density
\be\label{Pdx}
P(x)\dd x={\dd x\over\sqrt{4\pi Dt}}\,e^{-x^2/4Dt}~.
\ee
An important result of Einstein's computation was the computation of the 
diffusion constant $D$ in terms of the microscopic scales $l$ and $\t$.

The probability distribution \eq{discrete} satisfies an interesting property,
characteristic of birth processes \cite{karlinmcgregor1,karlinmcgregor2}.
If we make $P$ into an infinite matrix $P_{ij}(t)$, where $i$ and $j$ denote
positions along the $x$-direction, and $m$ the number of steps between the
state $i$ and the state $j$, then we can write the time evolution as follows:
\be\label{infiheat}
P'(t)=AP(t)~.
\ee
$A$ is an infinite matrix whose only non-zero entries are on the diagonal, and
one position above or below it, that is, only the matrix elements $A_{ii}$,
$A_{i,i+1}$ and $A_{i,i-1}$ are non-vanishing. This expresses the fact that
the particle moves one step at the time right or left, but cannot jump. The
intial condition is obviously:
\be
P(0)={\mbox{id}}~.
\ee
Notice that, if we take $\t\rightarrow0$, this results in the delta-function
shape of the continuous case.

Equation \eq{infiheat} is really the infinitesimal version of the heat equation
satisfied by the continuous density \eq{Pdx}. There is a nice reformulation of
this equation as a difference equation \cite{kac}. Writing $A_{ii}=-(\l_i+\m_i)$,
$A_{ij}=\l_i$ if $j=i+1$ and $\m_i$ if $j=i-1$, the matrix determines a system
of polynomials by means of the recurrence relations
\bea
Q_0(x)&=&1\nn
-xQ_0(x)&=&-(\l_0+\m_0)Q_0(x)+\l_0Q_1(x)\nn
-xQ_n(x)&=&\m_nQ_{n-1}(x)-(\l_n+\m_n)Q_n(x)+\l_nQ_{n+1}(x)~.
\eea
Karlin and McGregor further showed that for any system of this type there 
exists a positive measure $\psi$ for which the following orthogonal relations 
hold:
\be
\int_0^\infty\dd\psi(x)\,Q_i(x)Q_j(x)={\d_{ij}\over\pi_i}
\ee
where the moments are given by $\pi_0=1$, $\pi_n={\l_0\l_1\ldots\l_{n-1}\over
\m_1\m_2\ldots\m_n}$. Such a measure is called a solution of the Stieltjes
moment problem. Extending the range of integration to all reals, one can
construct solutions of the analogous Hamburger moment problem.

The probability matrix is now given in terms of these polynomials by 
\be
P_{ij}(t)=\pi_j\int_0^\infty\dd\psi(x)\,e^{-xt}Q_i(x)Q_j(x)~,
\ee
and an important result of Karlin and Mc Gregor is that the determinant
\be
P\left({\over}t;\psi;\{i\},\{j\}\right)=\det P_{ij}(t)
\ee
is positive definite. This extends to the continuous case as well, a property
that we will use later on.

\subsection{Vicious walkers and Brownian motion probabilities}

In this paper we consider Brownian motion of a particular type: the so-called
{\it vicious walkers} \cite{fisher}. Vicious walkers are movers whose 
trajectories are not allowed to intersect during the whole motion. Let us
however consider the case of harmless movers first. These perform Brownian
motion, but their trajectories are allowed to intersect at any time. For a
single mover moving on a line, we compute the probability distribution of 
going from poing $x$ to point $y$ in time $t$, given a diffusion coefficient
$D$ in the medium. This is given by
\be\label{ptxy}
p_t(x,y)={1\over\sqrt{4\pi Dt}}\,e^{-(x-y)^2/4Dt}~.
\ee
This Brownian motion probability has some elementary but important properties
\cite{bochner}. We will list them here for this simple case since they
generalize to higher dimensions and underlie the matrix
model description of Brownian motion and Chern-Simons.\\
\\
1) It is the unique kernel of the solutions of the heat equation with prescribed boundary conditions:
\be\label{heateq}
{\pa\over\pa t}\,p_t(x,y)=D\Delta p_t(x,y)~,
\ee
which is usually denoted by $K(x,y;t)$. From now on we will set $D=\half$.\\
\\
2) It can be obtained as the continuum limit of a discrete random walk, as we reviewed
in the previous section.\\
\\
3) It is a positive-definite quantity and symmetric under interchange of $x$ and $y$. The former property is trivial in the 
one-dimensional case but it is a non-trivial result due to Karlin and 
McGregor \cite{karlinmcgregor1,karlinmcgregor2} for non-intersecting paths in higher dimensions. 
This connects the 
probabilistic interpretation with the Stieltjes-Wigert polynomials that 
naturally appear in Chern-Simons theory \cite{miguel}, as we will see.\\
\\
4) Finally, $K(x,y;t)$, being a probability, has the extensivity property
\be\label{extens1d}
\int\dd\m(z)\, K(x,z;t)K(z,y;s)=K(x,y;t+s)~.
\ee
Here, the integral runs over all reals\footnote{In this, our conventions are
different from those in \cite{bochner}, where -- for the non-compact range --
only functions over the positive reals are considered. Further, the heat 
equation considered here corresponds to the case $\g=0$ in \cite{bochner},
in which case the Bessel function is simply a cosine and the integration
measure is simply $\dd\m(z)=\sqrt{2/\pi}\dd z$. Therefore, in the case 
considered here and with the conventions of \cite{bochner} the kernel would be 
$K(x,y;t)=1/\sqrt{t}\exp(-(x^2+y^2)/2t)\cos(ixy/t)$.}. This is sometimes called
the Chapman-Kolmogoroff equation, and is a consequence of the Markov property
(`lack of memory' of the process). This condition also completely fixes the 
normalization of $K$, which in the case \eq{ptxy} integrates to 1. In the 
generalization we will consider later on, the normalization is still fixed by the extensivity property, but it is not equal to 1 since we are dealing with 
conditional probabilities.

It is interesting to note that these expressions can be generalized along the lines of \cite{bochner} to include first order derivatives in $x$ in the heat equation. These roughly correspond to an external gravitational field, and the solutions are given in terms of Bessel functions. It would be interesting to see whether there is a gauge theory analog of such terms.

The generalization of the above for more than one particle is obvious:
\be\label{ptn}
p_{t,N}(x,y)={\frac{1}{(2\pi t)^{N/2}}}\,e^{-{\frac{|x-y|^{2}}{2t}}},
\ee
and $N$ is the number of particles. This is the case that the particles are
non-interacting, and it can equivalently be regarded as the
product of the probabilities of $N$ single movers on a line, i.e. the
probability for $N$ movers on a line to start at positions $y_{1},\ldots, y_{N}$ and end up at $x_1,\ldots ,x_N$ after time $t$.

From the point of view of the heat equation, however, \eq{ptn} is not quite
the higher-dimensional generalization of the heat kernel of \eq{heateq}, where
now the Laplacian would be an $N$-dimensional flat space Laplacian. The
fundamental solution is only obtained after we take the determinant, as we will
now explain.

\begin{figure}
\begin{center}
\includegraphics[height=2.5in,width=4in]{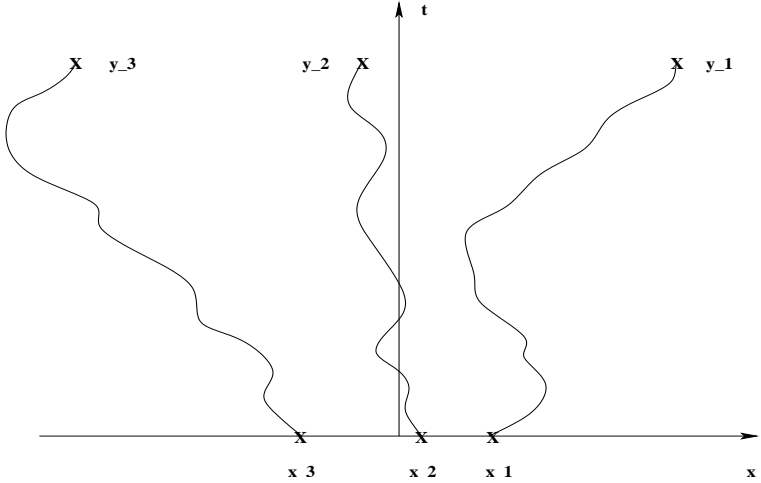}
\caption{\small Three vicious walkers performing Brownian motion from
$x_1,x_2,x_3$ to $y_1,y_2,y_3$.}\label{fig1}
\end{center}
\end{figure}

From the point of view of the heat equation we are naturally led to 
consider {\it vicious} walkers, that is, walkers whose trajectories do not
intersect at any time (see Fig. \ref{fig1}). The quantity that we are then interested in is the probability of going from a state $x_1,\ldots,x_N$ to a state $y_1,\ldots,y_N$ in time $t$ during which the particles perform free Brownian motion but are otherwise non-intersecting. That is, we do not enforce the non-intersecting condition by adding a force by hand, but consider rather those paths among all the possible paths such that the particles do not intersect. Their quotient then gives the desired probability density. Further, we label the particles such that at time $t=0$, $x_1>x_2>\ldots>x_N$. Since the trajectories are non-intersecting and the
particles are distinguishable, this condition remains true at all times.

It is this condition $x_1>x_2>\ldots x_N$ that allows us to think of the
motion in terms of a single particle moving in the fundamental Weyl chamber
of $U(N)$. Indeed, if $x_i$ now instead of labeling the position of a particle
on the line, labels the $i$th coordinate of a particle in an $N$-dimensional
space, then the motion is restricted to the fundamental Weyl chamber of 
$U(N)$. Later on we will develop this point of view further and see how it
generalizes to other groups.

The probability in the non-intersecting case is easy to compute thanks to the
fact that the heat equation is linear and we can superpose its solutions.
Indeed, we can use the method of images to write the number of non-intersecting
walks as the number of free walks, from which we appropriately subtract 
``bad'' walks. This is most easily done by thinking of this as motion of a
single particle in $N$ dimensions, in which case the ``bad'' walks are walks
which at some point hit one of the walls of the fundamental Weyl chamber. This
is explained in section \ref{generalformulation}.

Using the method of images (see also section \ref{generalformulation}), the probability is given by:
\bea\label{ptN}
p_{t,N}(\l,\m)&=&{1\over(2\pi t)^{N/2}}\,e^{-{|\lambda|^2 
+|\mu|^2\over2t}}\, \det(e^{\lambda_i\mu_j/t})_{1\leq i<j\leq N}\nn
&=&{1\over(2\pi t)^{N/2}}\det(e^{-{(\l_i-\m_j)^2\over2t}})_{1\leq i<j\leq N}~,
\eea
where we now labeled the initial and final posititions as two vectors $\l$ and 
$\m$ in ${\mathbb R}^N$.

It is easy to see that this probability vanishes, as it should, when any two
particles hit each other or, alternatively, when the particle moving in the 
fundamental Weyl chamber hits one of the walls. Indeed, in that case 
$\m_i=\m_{i+1}$ for some $i$, and the determinant vanishes identically.

From the representation in the last line of \eq{ptN} we immediately see that
the interacting probability distribution inherits from the free Brownian motion
case its homogeneity, that is, it is invariant under simultaneous constant
shifts $\l_i\rightarrow\l_i+c$, $\m_i\rightarrow\m_i+c$. In the next section
we will study some more properties of $p_{t,N}(\l,\m)$.

Let us briefly discuss how this interacting probability satisfies the requirements 1)-4). As for
1), consider the $N$-dimensional heat equation
\be
{\pa\over\pa t}f(\l_1,\ldots,\l_N)=\half \Delta f(\l_1,\ldots,\l_N)~.
\ee
We are interested in antisymmetric solutions of this equation, $f(\l_1,\ldots,\l_N)$, with boundary 
condition $f_0(\l)$ at $t=0$. The unique solution with these boundary conditions is \cite{itzzub}
\be
f(\l)={1\over N!}\prod_{i=1}^N\int_{{\mathbb R}^N}\dd\m_i\,K(\l,\m;t)\,f_0(\m)
\ee
and
\be
K(\l,\m;t)={1\over(2\pi t)^{N/2}}\,\det(e^{-(\l_i-\m_j)^2/2t})
\ee
is precisely the Brownian motion density $p_{t,N}(\l,\m)$.

Properties 2) and 3) were already discussed earlier, so we will
skip the details. Positivity of the kernel $K(x,y)=p_{t,N}(x,y)$ was proven in 
\cite{karlinmcgregor2}, and its symmetry with respect to the vectors $x,y$ is clear from the explicit 
expression.

The extensivity property deserves a few more comments. In this case it reads
\be\label{extensivity}
p_{t+s,N}(x,y)=\int_C\dd^Nz\,p_{t,N}(x,z)\,p_{s,N}(z,y)
\ee
as we prove in appendix \ref{extens}. This simply expresses the fact that the probability of going from $x$ to $y$ in time $t+s$ is the same as that of going from $x$ to $y$ in time $t$, times that of going from $y$ to $z$ in time $s$, where we integrate over all possible intermediate states $y$. Notice that the normalizations in \eq{ptN}, and in particular also the fact that we integrate over the fundamental Weyl chamber $C$ only and not over ${\mathbb R}^N$, are crucial for the proper normalization of \eq{extensivity}. By definition, $p_{t,N}(x,y)$ --being a conditional probability-- does not integrate to one, as not all possible paths from $x$ to $y$ are non-intersecting. 

Let us anticipate that \eq{extensivity} is equivalent to a matrix model computation of the partition function of Chern-Simons theory on $S^3$. Actually it is more general than that, providing also a computation of Wilson loops.

Notice that Fisher \cite{fisher} includes an additional ``framing'' factor:
\be
\ti p_{t,N}(\l,\m)=e^{-N\s t} {1\over(2\pi t)^{N/2}}\,e^{-{|\lambda|^2 
+|\mu|^2\over2t}}\, \det(e^{\lambda_i\mu_j/t})_{1\leq i<j\leq N}~.
\ee
This corresponds to the total partition function, and is related to the 
total weights of the probabilities in the lattice of the underlying discrete
random walk.

In the next few sections we show how these probabilities are related to 
Chern-Simons quantities.

\section{Chern-Simons theory as Brownian motion: the $U(N)$ case}
\label{CStheory}

\subsection{Chern-Simons theory and surgery}

In this subsection we briefly review some aspects of Chern-Simons theory in 
three dimensions that we will need in what follows. 

Recall that Chern-Simons theory is a quantum field theory whose action is 
built of a Chern-Simons term involving as gauge field a gauge connection 
associated to a group $G$ on a three-manifold $M$ \cite{witten}. The 
action is:
\begin{equation}\label{CSaction}
S(A)={k \over 4 \pi} \int_M {\rm Tr} \Bigl( A \wedge \dd A + {2 \over 3}
\, A\wedge A \wedge A \Bigr)~.
\end{equation}
$k$ is the coupling constant and the trace is taken in the fundamental
representation. This action is invariant under gauge 
transformations; however, in order for the theory to make sense under large
gauge transformations the parameter $k$ needs to be an integer. In the
partition function one then integrates over all possible gauge field 
configurations:
\be
Z=\int{\cal D}A\,e^{iS_{\tn{CS}}}~.
\ee

At large $k$, the action \eq{CSaction} gives the equation of motion $F=0$; 
classically, therefore, the Chern-Simons path integral is dominated by flat
connections.

Notice that in order to define the action \eq{CSaction} there is no need to
choose a metric; indeed, this theory is diffeomorphic invariant at the 
classical level, and depends only on the global properties of the manifold on
which it is defined. This remains true at the quantum level, up to a subtlety.
To evaluate the partition function, one needs to fix the gauge symmetry. In
doing so, a choice of metric is required. This by itself does not present a
problem; although the resulting terms are not topological invariant but 
depend on the choice of metric, one can add a counterterm (even though there
are no divergences involved) that requires a choice of metric, such that the 
total combination is a topological invariant. In doing so, however, one needs
to fix a choice of trivialization of the tangent bundle, in other words a 
framing of the manifold. Thus, the resulting theory is a topological invariant
of framed manifolds.

Non-trivial observables in Chern-Simons theory are Wilson loops:
\be\label{defWilson}
W_\l(C)={\mbox{Tr}}_\l{\cal P}\exp\int_C\,A_\m\dd x^\m~.
\ee
This Wilson loop is labeled by a representation $\l$ of the gauge group, and
a closed loop $C$ in $M$. This knot can have non-zero crossing number,
which is the minimal number of crossings needed when we draw a two-dimensional
picture of the knot (for an introduction to knot theory, see \cite{murasugi,lickorish}). Knots are topological invariants, in that 
they depend only on the topology of the manifold and on the topological class 
of the embedding of the loop in the manifold. Indeed, again one can see from
\eq{defWilson} that the definition is independent of the choice of metric. We
will later see how to compute Wilson loops in practice. 

It is clear that one can also consider expectation values of products of Wilson
loops on knots that consist of several links. In that case, every link comes with
its own representation $\l_i$, and one should sum over the different links in 
\eq{defWilson}. For knots
consisting of $n$ links, the associated invariant is
\be
W_{\l_1\ldots\l_n}=\bra W_{\l_1}(C_1)\ldots W_{\l_n}(C_n)\ket~.
\ee
This expression factorizes if the links are disjoint. The classic result of \cite{witten} is that these
quantum field theory expectation values can actually be computed exactly, and are
in fact given by certain polynomials (actually rational functions, when properly normalized)
in the exponentiated coupling and rank of the gauge group, reproducing known
knot invariants in the mathematical literature.

Let us ilustrate this topological invariance in the abelian case. Consider for
simplicity the case of two loops, $C_1$ and $C_2$. The equations of motion in
the presence of these sources take the form:
\be
\e^{\m\n\l}\pa_\n A_\l=-{4\pi\over k}\,J^\m~,
\ee
something reminiscent of Maxwell's equations in four dimensions, where the 
gauge potential is playing the role of the magnetic field and the source for
the Wilson loops is in this case
\be
J^\m(x)=e_1\int_0^{2\pi}\dd s\,{\dd y^\m\over\dd s}\,\d^{(3)}(x-y(s))
+e_2\int_0^{2\pi}\dd t\,{\dd z^\m\over\dd t}\,\d^{(3)}(x-z(t))
\ee
where $e_1$ and $e_2$ are the charges of the particles going around the loop,
and $y^\m(s)$ and $z^\m(t)$ are embeddings of the loops in ${\mathbb R}^3$. The
above can be easily solved giving the gauge potential as an integral around the
loop. In the classical limit we get:
\be
\bra e^{i\int\sm{d}^3 x\,J^\m A_\m}\ket =e^{{ie_1\over2}\oint_{C_1}A_\m
\sm{d} x^\m +{ie_2\over2}\oint_{C_2}A_\m\sm{d} x^\m}~,
\ee
and filling in the value of the gauge field this gives:
\be\label{braWket}
\bra W\ket=e^{-{2\pi i\over k}\sum_{a,b}e_ae_b\chi(C_a,C_b)}
\ee
and the sum runs over $a=1,2$. $\chi(C_1,C_2)$ is here the Gauss linking 
number:
\be
\chi(C_1,C_2)={1\over4\pi}\oint_{C_1}\dd x^\m\oint_{C_2}\dd y^\n\,\e_{\m\n\l}
{(x-y)^\l\over|x-y|^3}~.
\ee
As long as $C_1$ and $C_2$ do not intersect at any point, the Gauss linking number depends only on topology. It is given by:
\be
\chi(C_1,C_2)=\half\sum_p\e(p)~,
\ee
where $p$ are the points where (on a two-dimensional picture of the link) the
links cross, and we should not count self-intersections of the knots, which appear on projection of the
knot on the plane. $\e(p)$ is a sign assigned to the crossing, with a $+1$ for an upper pass and a $-1$
for an underpass. Notice that when the orientation of one of the loops is reversed, the overall sign 
changes.

The expectation value \eq{braWket} also contains two self-interaction terms where $a=b$ that we have not considered so far. The trouble with such a term is that, even though it is well-defined and finite, it is not a topological invariant. One obtains a topological invariant if one provides knots with a framing, that is, one chooses a normal vector field along the knot, or, in other words, one ``fattens'' the knots, making them to small strips, and defines the self-linking number to be the linking number of boths sides of the strip. This is really a point-splitting regularization. The final result does actually depend on the choice of framing; different choices of framing change the self-linking number by an integer, and in this way we can compare different framings. Although there are no divergences involved, this is analogous to a choice of counterterms in quantum field theory. This discussion actually extends to the non-abelian, full quantum case, and we will quote the result here for future reference. A Wilson loop with $n$ links in representations $\l_1,\ldots,\l_n$ changes according to:
\be
\bra W_{\l_1\ldots\l_n}\ket \rightarrow e^{2\pi i\sum_{i=1}^ns_ih(\l_i)} \bra W_{\l_1\ldots\l_n}\ket~,
\ee
where $h(\l)$ is the conformal weight of the representation $\l$ in the WZW model:
\be
h(\l)={C(\l)\over2(k+g)}~,
\ee
$C$ being the Casimir operator and $g$ the rank of the gauge group\footnote{For a summary of our group theory conventions, see appendix \ref{partitions}.}. $s_i$ is an integer, denoting the number of units by which we change the self-linking number of the $i$th knot, according to the counting of crossings described above.

Let us briefly recall canonical quantization of Chern-Simons \cite{witten}. 
This gives an alternative viewpoint to path integral techniques, and in 
particular it allows us to compute Chern-Simons quantities in more complicated 
manifolds. The idea is to construct a three-manifold $M$ by {\it surgery}, that is, by cutting the manifold into tractable pieces and then
gluing them together. Locally, the manifold looks like 
$\Sigma\times{\mathbb R}$, where we regard ${\mathbb R}$ as ``time'', and by 
fixing temporal gauge one easily sees that the Hilbert space associated to 
$\S$ is the Hilbert space of flat connections on $\S$ modulo gauge 
transformations. We refer to \cite{witten} for a discussion of this.

We will be interested in the cases where one can obtain $M$ by gluing together two solid tori
after performing some diffeomorphism. The Hilbert space associated with the boundary of each solid 
torus is then the Hilbert space of the WZW model 
\cite{witten}, which is the space of integrable representations of highest weight
at level $k$. In particular, the vacuum state, corresponding to the solid torus without insertions of 
Wilson lines, is represented by the Weyl vector $|\rho\ket$ (see appendix 
\ref{partitions}), which is the trivial representation or an empty Young tableau. Inserting a Wilson 
loop in representation $\l$ around the non-contractible cycle of the solid torus gives a state in the 
Hilbert space, $|\l\ket$, where $\l$ is integrable.

Let us consider how to get $S^3$ from this. To get $S^3$, we need to identify
the $A$-cycle of one of the tori with the $B$-cycle of the other one, and
viceversa. In order to do that, we act with the $SL(2,{\mathbb{Z}})$ generators $S$ and $T$:
\bea
T&=&\left(
\begin{array}{cc}
1 & \,\,\,1 \\
0 & \,\,\,1
\end{array}
\right)~,\nn
S&=&\left(
\begin{array}{cc}
0 & -1 \\
1 & \,\,\,0
\end{array}
\right)~,
\eea
where $T$ acts on the complex structure of the torus as $\t\rightarrow\t+1$, that is, it shifts the 
$A$-cycle by a $B$-cycle, and  $S$ acts as $\t\rightarrow-1/\t$, i.e., it sends $A$ to $-B$ and $B$
to $A$. From the latter operation we clearly get an $S^3$ after gluing both tori. Thus, we have to 
compute the matrix element
\be\label{S00}
Z_{\sm{CS}}(S^3)=\bra\r|S|\r\ket=S_{00}~.
\ee
That is, we need to represent $S$ on the states $|\l\ket$. This is the well-known modular matrix
of the WZW model, which we quote here together with the other generator $T$ for future reference:
\bea\label{ST}
S_{\l\m} &=&{\frac{i^{|\Delta _{+}|}}{(k+g)^{r/2}}}\,|P/Q^{\vee}|^{-{\frac{1}{2}}\,}
\sum_{w\in W}\epsilon (w)e^{-{2\pi i\over k+g}(\lambda ,w\cdot \mu )}\nn
T_{\l\m} &=&\delta _{\lambda \mu }\,e^{{\frac{2\pi iC(\lambda )}{%
2(k+g)}}-{\frac{2\pi ic}{24}}}~.
\eea
The central charge is $c=k\,{\mbox{dim}}\,g/(k+g)$, $C(\lambda )$ is the Casimir
of the representation $\lambda $, $\Delta _{+}$ is the set of positive
roots, $P$ is the weight lattice, and $Q^{\vee }$ is the coroot lattice.

In the gluing we could have acted with additional external factors of $T$. However, this only
amounts to a change of framing of the $S^3$, as discussed earlier. We now see that framing factors are
proportional to $T$ as in \eq{ST}. In what follows, we construct $S^3$ with the transformation $TST$.

By considering more general $SL(2,{\mathbb{Z}})$ transformations, we can construct more generic 
manifolds such as lens spaces and Seifert manifolds. The most general manifold that we can obtain from
gluing of two solid tori (that is, by doing surgery on an unknot inside $S^3$) is by applying the
$SL(2,{\mathbb{Z}})$ matrix
\bea
U&=&\left(
\begin{array}{cc}
p & \,\,\,r \\
q & \,\,\,s
\end{array}
\right)~,
\eea
where $ps-qr=1$. The representation of this matrix on affine characters is 
\cite{jeffrey,rozansky,marcos}:
\be\label{U}
U^{(p,q)}_{\l\m}={(i\,{\mbox{sgn}}\,q)^{|\D_+|}\over(l|q|)^{r/2}}\,e^{-{id\pi\over12}\Phi(U^{(p,q)})}
\,|P/Q|^{1/2}\sum_{n\in Q^\vee/qQ^\vee}\sum_{w\in W}\e(w)e^{{i\pi\over lq}[p\l^2-2\l(ln+w\m)
+s(ln+w\m)^2]}
\ee
where $Q$ is the root lattice, $\F(U^{(p,q)})$ is the Rademacher function
\be
\F\left[{p\atop q}{r\atop s}\right]={p+s\over q}-12s(p,q)~,
\ee
and $s(p,q)$ is the Dedekind sum
\be
s(p,q)={1\over4q}\sum_{n=1}^{q-1}\cot{\pi n\over q}\cot{\pi np\over q}~.
\ee
In section \ref{generalformulation} we will work out in detail the case of the lens space 
$S^3/{\mathbb{Z}}_p$. More general manifolds can be obtained by surgery on more general knots. See for example \cite{jeffrey,rozansky,marcos} for Seifert spaces.

Finally, we give the explicit expressions for the partition function and Wilson loops of
Chern-Simons in $S^3$, which we will use later. The partition function on $S^3$ for gauge 
group $U(N)$ is given by working out \eq{S00}:
\be\label{partitiefunctie}
Z_{\sm{CS}}(S^3)={1\over(k+N)^{N/2}}\prod_{j=1}^{N-1}(2\sin{\pi j\over k+N})^{N-j}.
\ee
The unnormalized expectation value of a Wilson loop around the unknot $C$ is:
\be\label{unknot0}
W_\l(C)={1\over(k+N)^{N/2}}\prod_{i<j}2\sin{\pi(\l_i-\l_j)\over k+N}~.
\ee

\subsection{Chern-Simons on $S^3$: the probability of reunion}

In this section we work out in detail the relation between Brownian motion
and Chern-Simons theory for the case of the partition function, which
corresponds to the probability of reunion of random walkers.

We will now evaluate the above probability density in the simplest case: the
probablity of reunion, and see how this gives us the partition function of
Chern-Simons on $S^3$ with gauge group $U(N)$. The probability of reunion is 
defined \cite{fisher} as the probability of, given that we start with a 
configuration where all the movers are equally spaced,
\be\label{bc0}
\l_j=(c-j)a
\ee
ending up with the same configuration after time $t$: $\m_j=\l_j$. The special boundary condition 
\eq{bc0} we will call $\l_{0j}$. Notice that $a$ is the initial
spacing between two movers:
\be
\l_{0j+1}-\l_{0j}=a
\ee
and at this point it can be positive or negative. Since we arrange our 
coordinates such that $\l_1>\l_2>\ldots\l_N$, we will take $a$ to be positive. 
Let us here emphasize that, even though in the discrete case $a$ would 
naturally have an interpretation as a lattice length and hence a minimal 
distance between two movers, that is not the case for Brownian motion, where 
two particles can be arbitrarily close to each other as long as they do not 
intersect. Notice that to get the trivial representation, labeled by the Weyl
vector, we have to take (see appendix \ref{partitions})
\be
\l_{0j}=\r_j={N+1\over2}-j
\ee
which in particular satisfies $\sum_{i=1}^N\r_i=0$, a property that we will
use later. However, as mentioned earlier, the overall constant is irrelevant as
it only corresponds to a constant shift. In this and the next subsection we do not keep track of 
coupling-independent phase factors.

In appendix \ref{vdmonde} we show that the probability of reunion is given
by
\bea\label{prob00}
p_{t,N}(\l_{0},\l_{0})&=&{1\over(2\pi t)^{N/2}}\,
\prod_{k=1}^N(1-q^k)^{N-k}\nn
&=&{1\over(2\pi t)^{N/2}}\,q^{{1\over6}N(N^2-1)}
\prod_{k=1}^N(1-q^{-k})^{N-k}~,
\eea
where $q=e^{-a^2/t}$. If we compare this with \eq{partitiefunctie} (see also \eq{mm2}), 
we see that up to 
normalizations both expressions are the same\footnote{Notice that at this point
the relation $-1/t=g_s$ could involve a plus sign instead of a minus sign. This
would result in a different choice of framing, which is therefore immaterial
for the interpretation as the partition function of Chern-Simons on $S^3$. We
will see, however, that this arbitrariness is fixed by comparing our 
expressions with the WZW modular $S$-matrix, which will give the minus sign. Notice, however, that ultimately this is irrelevant as all expressions are holomorphic in the coupling. In particular, the relation $1/t=g_s$ should work as well if we identify the Brownian motion probability with the complex conjugate of $S$.} if we set $a^2=1$ and
\be\label{coupling}
-{1\over t}=g_s={2\pi i\over k+N}~.
\ee
Then we get
\be
Z_{\sm{mm}}= p_{t,N}(\l_0,\l_0)
\ee
and in canonical framing
\be\label{partfunction}
Z_{\sm{CS}}(S^3)=e^{-{1\over12}g_sN(N^2-1)}
p_{t,N}(\l_0,\l_0)~.
\ee

This is the basic result in \cite{brownian} concerning the relation between
the partition function of Chern-Simons on $S^3$ and the probability of
reunion of $N$ vicious Brownian movers.

\subsection{Wilson lines: the unknot and the Hopf link}

The next non-trivial case is to consider Brownian motion where we fix the
initial positions to be equally spaced, but leave the final positions 
arbitrary. Thus we fix
\be
\m=\r a
\ee
where $\r$ is the Weyl vector (see appendix \ref{partitions}) and leave $\l$ 
arbitrary. The resulting determinant \eq{ptN} can now be 
readily computed. In a first step we use the standard Vandermonde determinant
formula \eq{Vdm}. In a second step, we use the identity \eq{det2}. The
result is:
\be\label{pl0}
\det(e^{-(\l_i-\r_ja)^2/2t})=e^{-{|\l|^2\over2t}-{1\over24t}N(N^2-1)}
\prod_{i<j}2\sinh\left({\l_i-\l_j\over2t}a\right)~.
\ee
Of course, we have simply proved the Weyl denominator formula for the symmetric
group. As a simple check, filling in $\l=\r$ gives us
\be
p_{t,N}(\r,\r)={1\over(2\pi t)^{N/2}}\,e^{-{1\over12t}N(N^2-1)}\prod_{i<j}
2\sinh({j-i\over2t})~,
\ee
which reduces to \eq{ptN} after working out the product and filling in the 
value of $t$.

What quantity could the more general equation \eq{pl0} possibly be in 
Chern-Simons theory? In fact it is nothing but the expectation value of a Wilson loop on $S^3$ in a representation $\l$ of $U(N)$, as we can see from \eq{unknot0}. This Wilson loop winds an unknot, which in $S^3$ is the unique knot with one 
link component since there are no non-trivial cycles. The Wilson loop is 
however not in canonical framing. To see this, it is easiest to write
\be\label{unknot}
p_{t,N}(\l,\r)= e^{{2\pi i\over12}N(N^2-1)}\,\bra0|TST|\l\ket
\ee
where the modular matrices $T$ and $S$ are given as in \eq{ST}. In appendix \ref{framing} we 
will analyze this framing in detail.

Having written down the representation \eq{unknot}, it is now straightforward
to generalize this to the probability density with arbitrary initial and
final states: we simply compute the operator $TST$ within arbitrary external
states $\l$ and $\m$, corresponding to the two boundary contitions:
\be
p_{t,N}(\l,\m)= e^{{2\pi i\over12}N(N^2-1)}\,\bra\m|TST|\l\ket~,
\ee
as is easily checked.

\section{Brownian motion with a wall: $SO(N)$ and $Sp(N)$ Chern-Simons}

An interesting generalization of the Brownian movers on a line is to see what happens 
when we impose an additional restriction by inserting a wall at the origin, $\l=0$ \cite{grabiner}. Thus
we consider Brownian motion of $N$ non-intersecting movers on the half-line 
${\mathbb R}_+$, that is, the coordinates of the movers now satisfy 
$\l_1>\l_2>\ldots>\l_N>0$. We immediately see that this is the Weyl chamber of $SO$ and $Sp$, and we will
return to this interpretation momentarily.

Using the method of images, we can again compute this probability from the free case \eq{ptxy}. Namely, on top of the reflections about the forbidden walls $\l_i=\l_j$,
$i\not=j$, we need to reflect around the origin. We get 
\be\label{sosp}
p_{t,N}(\l,\m)=\sum_{\s\in S_N}\e(\s)\sum_{\ve_1\ldots\ve_N=\pm1}(-1)^{\ve_1+\ldots+\ve_N}
\prod_{i=1}^Np^0_t(\ve_i\l_{\s(i)}-\m_i)
\ee
where $p^0_t(\l,\m)={1\over\sqrt{2\pi t}}\,e^{-(\l-\m)^2/2t}$ is the free Brownian motion
probability for one particle. This can be rewritten as:
\bea\label{sosp1}
p_{t,N}(\l,\m)&=&{1\over(2\pi t)^{N/2}}\,e^{-{|\l|^2+|\m|^2\over2t}}
\sum_{\s\in S_N}\e(\s)\sum_{\ve_1\ldots\ve_N=\pm1}(-1)^{\ve_1+\ldots+\ve_N}
e^{\sum_{i=1}^N\ve_i\l_{\s(i)}\m_i/t}\nn
&=&{1\over(2\pi t)^{N/2}}\,e^{-{|\l|^2+|\m|^2\over2t}}
\sum_{\s\in S_N}\e(\s) \prod_{i=1}^N(e^{\l_{\s(i)}\m_i/t}-e^{-\l_{\s(i)}\m_i/t})\nn
&=&{1\over(2\pi t)^{N/2}}\,e^{-{|\l|^2+|\m|^2\over2t}}\,
\det(e^{\l_i\m_j/t}-e^{-\l_i\m_j/t})_{1\leq i<j\leq N}~.
\eea

We now come to the connection with the $SO(N)$ and $Sp(N)$ groups. \eq{sosp} is really
a sum over the Weyl chamber of $B_N$. Notice that the Weyl chambers of $B_N$ ($SO(2N+1)$) 
and $C_N$ ($Sp(2N)$) are the same, and indeed we can interpret \eq{sosp1} as Brownian
motion probabilities of a single mover moving from $\l$ to $\m$ in the Weyl chamber of 
$SO(2N+1)$ or $Sp(2N)$.

Asymptotic expressions for \eq{sosp1} were given in \cite{grabiner} at large $t$ for the case
of initial equal spacing condition. However, we can compute \eq{sosp1} rather explicitly 
using the Weyl denominator formula. Remember that the equal spacing condition amounts to
setting $\l$ to be the Weyl vector, $\l=\r$. Explicitly,
\bea
\r&=&\sum_{i=1}^N(N+1-i)e_i \,\,\,\,\,\,\,\,\,\,\,\,\,\,\,\,\,\,{\mbox{for }}Sp(2N)\nn
\r&=&\sum_{i=1}^N(N+\half-i)e_i \,\,\,\,\,\,\,\,\,\,\,\,\,\,\,\,\,\,{\mbox{for }}Sp(2N)~.
\eea
Thus, $\r_{\tn{Sp}}=\r_{\tn{SO}}+\half(1,1,\ldots,1)$. Obviously, although the Weyl 
chambers are the same, their root systems are different. In particular, some of the
positive roots differ by a factor of 2. That gives the factor of 1/2 difference in the 
above formulas for the Weyl vectors. Now setting
the initial condition $\l=\r$ equal to the Weyl vector, both
for $SO$ and $Sp$ groups we can rewrite \eq{sosp1} as:
\bea\label{sosp2}
p_{t,N}(\r,\m)&=&{1\over(2\pi t)^{N/2}}\,e^{-{|\r|^2+|\m|^2\over2t}}
\prod_{\a>0}(e^{(\a,\m)/2t}-e^{-(\a,\m)/2t})\nn
&=&{1\over(2\pi t)^{N/2}}\,e^{-|\r|^2+|\m|^2\over2t}2^{|\D|_+}\prod_{\a>0}\sinh
{(\a,\m)\over2t}~.
\eea
We now have to specify explicitly what the positive roots $\a_{ij}$ are. For $SO(2N+1)$ 
they are:
\bea
&&\{e_i-e_j\}_{i<j}\nn
&&\{e_i+e_j\}_{i<j}\nn
&&\{e_i\}~,
\eea
with $i,j=1,\ldots,N$. Thus we get:
\bea\label{so}
p_{t,N}(\r,\m)&=&{1\over(2\pi t)^{N/2}}\,e^{-{|\r|^2+|\m|^2\over2t}}2^{|\D|_+}
\prod_{i<j}\sinh{\m_i-\m_j\over2t}\sinh{\m_i+\m_j\over2t}\prod_{k=1}^N\sinh{\m_i\over2t}~.
\eea
By working with the determinantal expression directly rather than using the Weyl formula,
the two first $\sinh$ can be combined into a single Vandermonde term \cite{fultonharris}.
Having written the expression for generic $\m$, we now specialize to $\m=\r$ also, so 
that we can compare with the partition function of Chern-Simons. We get:
\be\label{sosp3}
p_{t,N}(\r,\r)={1\over(2\pi t)^{N/2}}\,e^{-|\r|^2/t}2^{|\D|_+}\prod_{k=1}^{2N+1}
\left(\sinh{k\over2t}\right)^{f(k)}
\ee
where $f(k)$ is given in \eq{fso1} of appendix \ref{SOSpCS}. In fact, the above is also the answer for $Sp(2N)$. In
that case, the positive roots are
\bea
&&\{e_i-e_j\}_{i<j}\nn
&&\{e_i+e_j\}_{i<j}\nn
&&2\{e_i\}~.
\eea
Thus we get the expression \eq{so} with an additional factor of 2 in the last term. Again,
specializing to $\m=\r_{\tn{Sp}}$, we get \eq{sosp3} with $f(k)$ as in formula \eq{fsp} of appendix \ref{SOSpCS}. 

Let us now compare with the partition function of Chern-Simons on $S^3$. It is equal to:
\be\label{socs}
Z_{\sm{CS}}(S^3)=S_{00}=(k+g)^{-1/2}\prod_{\a>0}2\sin{\pi(\a,\r)\over
k+g}
\ee
Up to the usual framing factor and roots of unity, this is precisely formula 
\eq{sosp2} with 
boundary condition $\m=\r$. In particular, equation \eq{sosp3} precisely agrees with the
expressions in \cite{sinhavafa} for the $SO(2N+1)$ and $Sp(2N)$ partition function of 
Chern-Simons on $S^3$. The identification between the coupling and the level is given by
\eq{coupling}, as before. The framing factor $e^{-|\r|^2/t}$ is given by 
$|\r|^2={1\over12}N(4N^2-1)$ for $SO(2N+1)$, and $|\r|^2={1\over6}N(N+1)(2N+1)$ for
$Sp(2N)$.

It should now also be clear that the more general expressions $p_{t,N}(\r,\m)$ and
$p_{t,N}(\l,\m)$ correspond to the expectation values of the unknot and the Hopf link,
respectively.

\begin{figure}
\begin{center}
\includegraphics[height=2.5in,width=4in]{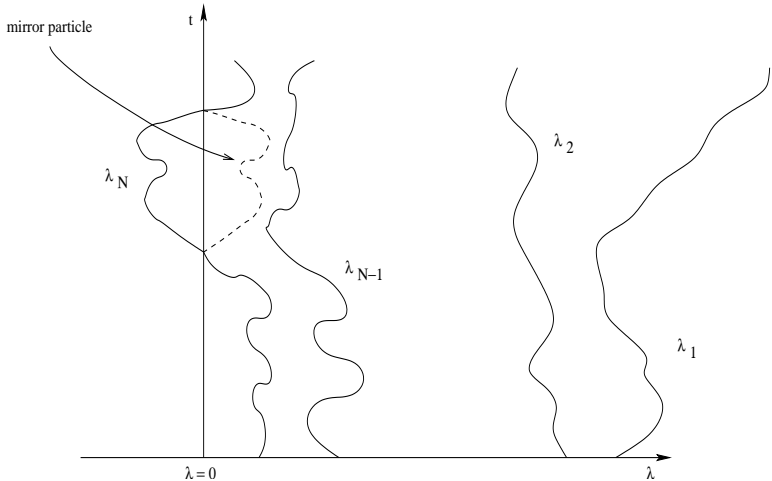}
\caption{\small Two-dimensional representation of motion in the Weyl chamber of 
$SO(2N+1)$ and $Sp(2N)$.}\label{fig2}
\end{center}
\end{figure}

Having done $SO(2N+1)$ and $Sp(2N)$, whose root systems are given by $B_N$ and $C_N$,
respectively, it is now natural to look at $SO(2N)$, which corresponds to $D_N$. The
Weyl chamber is now given by $\l_1>\l_2>\ldots\l_N$, and in addition $\l_{N-1}>-\l_N$.
This means that $\l_{N-1}>|\l_N|$ independently of the sign of $\l_N$. We can interpret
this as Brownian motion of $N$ particles on a line with a wall at $\l_N=0$ such that,
when the $N$th particle crosses it, a mirror particle traveling along the
mirror trajectory $|\l_N|$ is emitted, such that $\l_{N-1}$ is not allowed to intersect the trajectory
of the mirror particle (see Fig. \ref{fig2}). Notice that such paths are counted twice,
once from the particle that goes to negative $\l_N$, and once from the one that stays
at $\l_N\geq0$.

The positive roots of $SO(2N)$ are:
\be
\a_{ij}^{\pm}=e_i\pm e_j
\ee
with $1\leq i<j\leq N$. The Weyl vector is:
\be\label{weylso2n}
\r=\sum_{i=1}^N(N-i)e_i~.
\ee
The probability is now easy to work out and goes as before:
\be
p_{t,N}(\l,\m)=\half{1\over(2\pi t)^{N/2}}e^{-{|\l|^2+|\m|^2\over2t}}[\det(e^{\l_i\m_j/t}
+e^{-\l_i\m_j/t})+\det(e^{\l_i\m_j/t}-e^{\l_i\m_j/t})]~.
\ee
Notice that the last term is precisely the probability for $B_N$ and $C_N$. This term vanishes if some $\m_i=0$. By construction this was never the case for $B_N$ 
and $C_N$, since the Weyl chamber precisely satisfied the condition $\l_1>\l_2>\ldots>\l_N
>0$. However, for $D_N$ this is possible for the $N$th particle, which can indeed reach
$\l_N=0$ as depicted in Fig. \ref{fig2}. In fact, from \eq{weylso2n} we see that the Weyl
vector does have a zero. Thus in computing values of $p_{t,N}(\l,\m)$ with either $\l$ or
$\m$ equal to $\r$, the second term drops out. We can now expand the determinant for $\m=\r$ to get:
\be
p_{t,N}(\l,\r)={1\over(2\pi t)^{N/2}}\,e^{-{|\l|^2+|\r|^2\over2t}} \prod_{i<j} 4\sinh{\l_i-\l_j\over2}
\sinh{\l_i+\l_j\over2}~,
\ee
which is the expected result for the expectation value of a Wilson loop for the group $SO(2N)$. Filling in also $\l=\r$, what we get is the expression for the partition function of $SO(2N)$ Chern-Simons on $S^3$ \eq{socs}, as expected. The explicit expression is given in appendix \ref{SOSpCS}.

A further interesting case to analyze would be that of particles moving between two walls.

\section{Lie algebra wanderers: general formulation}\label{generalformulation}

We can generalize the above to any Weyl group associated to a root system \cite{gesselzeilberger}. In
particular, the root system can be finite (A-G series) or affine. The affine case was worked out in detail in 
\cite{sdh} and it was shown to be equivalent to 2dYM on the cylinder.

Thus we consider Brownian motion of a particle in the Weyl chamber $C$ of some root system $\D$. The Weyl 
group, $W$, generates ${\mathbb{R}}^r$ from the action on $C$, and $r$ is the rank. Let $\l$ and $\m$ be two points
in ${\mathbb{R}}^r$ (which, as usual, in the finite case will be chosen such as to correspond to 
representations of the corresponding group, and in the affine case will correspond to the integrable 
representations only). We can compute the conditional probability of going from $\l$ to $\m$ in time 
$t$ without leaving the fundamental Weyl chamber by the method of images, as follows. The argument is 
basically a continuous version of a well-known argument by Zeilberger (see \cite{gesselzeilberger} and
references therein). We denote the total number of 
walks\footnote{This is actually a probability, but in this section we will continue to talk of the number 
of paths; of course, the difference is just divinding out an infinite factor.} from $\l$ to $\m$ in time 
$t$ by $p^0_t(\l,\m)$. Notice that $p^0_t(\l,\m)$ is the free Brownian motion probability,
\be
p^0_{t,r}(\l,\m)={1\over(2\pi t)^{r/2}}\,e^{|\l-\m|^2/2t}~,
\ee
where $r$ is the rank of the gauge group. These are walks in ${\mathbb{R}}^r$. To count the number of
walks that stay within $C$, we divide all the walks in $p^0_t(\l,\m)$ into good and bad waks, according to
whether they do or do not remain within the Weyl chamber. We will call these $p_{t,r}(\l,\m)$ and 
$q_t(\l,\m)$, respectively, and we are obviously interested in computing $p_{t,r}(\l,\m)$. It is clear 
that we have:
\be
p_{t,r}(\l,\m)=p^0_{t,r}(\l,\m)-q_{t,r}(\l,\m)~.
\ee
The key step is now to use the method of images to show that the bad walks satisfy
\be\label{badwalks}
\sum_{w\in W}\e(w)\,q_{t,r}(w(\l),\m)=0~.
\ee
That is, for each bad walk from $w(\l)$ to $\m$ we can construct a mirror walk by reflecting with respect 
to the wall that was crossed last. Let us call the root that this wall corresponds to $\a\in\D$. The
wall is by definition perpendicular to $\a$, $(x,\a)=0$. The mirror walk in question is then another bad 
walk, constructed by taking the mirror part of the walk from $w_\a w(\l)$ to the crossing point, and 
from there continuing to $\m$ along the original walk. This gives a unique pairing between walks in the 
sum \eq{badwalks}. Since $w$ and $w_\a w$ have different sign, all such pairs cancel out. 

Now it is also clear that
\be
p_{t,r}(w(\l),\m)=q_{t,r}(w(\l),\m)\,\,\,\,{\mbox{if}}\,\,w\not={\mbox{id}}~,
\ee
since $w(\l)$ always lies outside $C$. Combining the above three formulas, we get

\be
p_t(\l,\m)=\sum_{w\in W}\e(w)p^0_t(w(\l),\m)={1\over(2\pi t)^{r/2}}
\sum_{w\in W}\e(w)e^{-|\l-w\m|^2/2t}~.
\ee
This is the desired generalization of the non-intersecting probability, and we should note that it holds both if $W$ is finite or affine \cite{sdh}.

We would next like to compute it in some natural ``vacuum'' state and compare it with Chern-Simons theory.
We will do this for the simple algebras. We take $\m=\r$ where $\r$ is the Weyl vector labeling the 
trivial representation. Using the Weyl denominator formula, we get:
\be\label{unknott}
p_{t,r}(\lambda ,\rho )={\frac{1}{(2\pi t)^{r/2}}}\,
e^{-{\frac{|\lambda|^{2}+|\rho |^{2}}{2t}}}
\prod_{\alpha >0}2\sinh {\frac{(\alpha ,\lambda)}{2t}}~,
\ee
where $\alpha $ are the positive roots. Under the identification $-1/t=2\pi i/(k+g)$, this expression is 
the (unnormalized) expectation value of a Wilson loop around the unknot. 

The partition function is obtained by setting $\l=\r$:
\be
Z_{\sm{CS}}(S^3)={(-i)^{r/2}\over(k+g)^{r/2}}e^{-{2\pi i\over 12(k+g)}\,g\,{\sm{dim}}\,g}
\prod_{\a>0}2\sin{\pi(\a,\r)\over k+g}
\ee
where we used the Freudenthal-de Vries strange formula $|\r|^2={g\over12}\,{\mbox{dim}}\,g$.

Notice that normalizing $p_{t,r}(\l,\r)$ by $p_{t,r}(\r,\r)$ gives the quantum dimension:
\be
\prod_{\a>0}{[(\l,\a)]\over[(\r,\a)]}~.
\ee

We can also rewrite the above directly in terms of the WZW modular matrices. Using the definitions
\eq{ST} and the value of the central charge $c=k\,{\mbox{dim}}\,g/(k+g)$, we get
\begin{equation}
p_{t,r}(\l,\m)=e^{{2\pi i\over12}\,{\sm{dim}}\,g}\,
(TST)_{\lambda \mu }~.
\end{equation}

Obviously, $S$ itself corresponds to the interaction term in the Brownian motion probability, 
and $T$ is the Boltzmann factor, which can be seen as a propagator in the diffusive
medium for translations over time $t$. It is now also clear that $p_{t,r}(\l,\m)$ itself gives
the (unnormalized) expectation value of the Hopf link invariant with representations 
$\l$ and $\m$.

In this language it is now much more clear why we are getting the partition function of 
Chern-Simons on $S^3$, and we can give an intuitive physical picture in terms of surgery that we 
will explain in more detail later. The boundary conditions $\l$ and $\m$ correspond to states in
the Chern-Simons Hilbert space associated with two solid tori. That is, we have two solid tori where
the cycle that is not filled in carries a representation $\l$, $\m$, respectively, of the gauge
group $G$. The probability tells us to propagate a random walker from $\l$ to $\m$ but performing 
a transformation $TST$, that is, we glue the two solid tori with a modular transformation 
$\t\rightarrow\t/(\t+1)$ before gluing. This gives us an $S^3$ in a framing that is not the canonical
one discussed in section \ref{CStheory}, but is the matrix model framing, which is the natural framing in
the mirror B-model on the resolved conifold, as we will see later \cite{akmv}. We will get back to this in section 
\ref{topstring}. We discuss other framings in the appendix \ref{framing}.

The matrix model will be analyzed in great detail in section \ref{matrixmodel}.

\subsection{Lens spaces and q-deformed 2d Yang-Mills}

We now discuss the case of the lens space $S^3/{\mathbb{Z}}_p$. This can be obtained by taking the
affine Lie algebra above, with an important subtlety that was discussed in \cite{sdh}. 

The affine Weyl chamber is 
\be
\ti W=W\ltimes T~,
\ee
where $T$ denotes translations in the coroot lattice $Q^\vee$. The Brownian motion probability in this fundamental Weyl chamber is then
\be\label{qtr}
q_{t,r}(\l,\m)={1\over(2\pi t)^{r/2}}\,\sum_{\g\in lQ^\vee}\sum_{w\in W}
\e(w)e^{-{1\over2t}|\g+\l-w\m|^2}~.
\ee
For real positive $t$, this is well-known to converge; that can for example be shown by rewriting it in
terms of $\Theta$-functions \cite{sdh}. When $1/t$ is $2\pi i$ times a rational number, however, as is
the case if we want to compare with Chern-Simons theory, the sum is not well-defined because the 
exponential is periodic. As explained in \cite{sdh}, a finite expression is obtained if one mods out by
this periodicity. Alternatively, one may regularize $-1/t={2\pi ip/l}+\e$, where $p$ and $l$ are 
integers and $\e$ real positive, and factor out an overall divergence, which goes like $\e^{-r/2}$. This is useful if one wants
to compare Chern-Simons on lens spaces and 2d Yang-Mills on the cylinder. Indeed, the appropriately
normalized \eq{qtr} gives the partition function of two-dimensional Yang-Mills theory on the cylinder
with the corresponding gauge group, where $\l$ and $\m$ label the holonomies of the gauge field at the
two ends of the cylinder. More precisely \cite{sdh}:
\be\label{cylinder}
Z_{\sm{2dYM}}(g,g';t)={(-il)^r|P/Q^\vee|\over S_{0\l'}S_{0\m'}}
\,q_{t,r}(\l,\m)~,
\ee
where we used the fact that the partition function is a class function of $g$ and $g'$ to conjugate
them into $e^{\l'}$ and $e^{-\m'}$, respectively, and we normalized $\l'=2\pi\l/l$, $\m'=2\pi i\m/l$. The
above was used to find the modular transformation properties of the partition function of 2d Yang-Mills on 
the cylinder \cite{sdh}. It was further shown that:
\be
{(ST^pS)_{\l\m}\over S_{\l\r}S_{\r\m}}=\left(p^2\e\over\pi\right)^{r/2}\,Z_{\sm{2dYM}}(g,g';t')~,
\ee
up to a phase, and $t'=(2\pi i/l)^2t$. This generalized previous relations between 2dYM and WZW matrices
\cite{witten2d,blauthompson}, to the case where the coupling times the area is $2\pi i$ times a rational 
number. As discussed in \cite{sdh}, the different normalizations are just what one expects, and the
overall divergence is obviously related to the fact that the Hilbert space is infinite, while we are
imposing the periodicity mentioned before. This is however only an overall divergence that can be 
renormalized.

One could however ask whether it makes sense to keep $t$ real and still talk about modular transformation
matrices. In fact, in \cite{aosv} it was shown that this is possible, and defines a quantum
deformed 2d Yang-Mills which is defined for values of $g_s$ which are not $2\pi i$ times a rational
number\footnote{This theory computes A-model amplitudes on a Calabi-Yau which is a sum of line bundles over a Riemann surface, where the Riemann surface is interpreted as the Riemann surface where the quantum deformed 2d Yang-Mills lives.}. As shown in \cite{aosv}, in that case the Hilbert space is infinite and one then typically sums 
over all representations and not only the integrable ones. The theory in question is the BF-theory
introduced in \cite{blauthompson}, where the scalar $\f$ (the $B$ of BF) is compact. The partition
function was shown to be:
\be
Z_{\sm{qYM}}(\S_g)=\sum_\l\left(S_{\l 0}\over S_{00}\right)^{2-2g}q^{pC_2(\l)/2}e^{i\theta\l}~.
\ee
In \cite{aosv} this was called quantum deformed 2dYM because $S_{\l 0}$ are the quantum dimensions rather
than the ordinary dimensions which appear in 2dYM. Let us consider the case of the sphere, $g=0$, with $\th=0$. In
this case, this is just a particular case of the partition function of ordinary 2dYM on the cylinder where $\l=\m=\r$,
and so the results of \cite{sdh} relate this to $q_{t,r}(\r,\r)$, \eq{qtr}. Such an expression was also found in 
\cite{aosv}. In \cite{sdh} it was also noticed that 2dYM on the cylinder and the BF theory of Blau
and Thompson are related, and from the above we see that indeed 
\be
Z_{\sm{2dYM}}(\r,\r)=Z_{\sm{qYM}}(S^2)~.
\ee
Notice that this holds for any value of $p$. The modular transformation properties of this theory were worked out in \cite{sdh} and we can now 
obtain from them the modular transformation properties of the corresponding A-model topological string amplitudes.  It would be worth to study 
this in more detail.
Notice also that an explicit expression for $Z_{\sm{2dYM}}(\l,\m)$ with its modular transformation properties are also available \cite{sdh}. It would be interesting to see if these can be interpreted as insertions on the sphere on the q-deformed 2dYM side.

Finally, let us remark that it is possible to obtain matrix model expressions for quantum deformed
2d Yang-Mills, which are useful to compute the large $N$ expansion. In some simple cases, it is not
hard to see that these are the same Chern-Simons matrix model expressions, analytically continued in
the coupling. We will get back to this issue in the near future \cite{sdhmt}.

\section{Brownian particles as mirror B-model branes}\label{topstring}

\subsection{The open topological string side}

Chern-Simons on $S^3$ describes A-model open topological strings on the deformed conifold $T^*S^3$:
\be
xu+yv=\m~.
\ee
In this section we briefly recall some aspects of the construction that we will use later on
(see, for example, \cite{akmv}).

It will be useful to recall the canonical quantization of Chern-Simons on $S^3$ \cite{emss,akmv}.
The Hilbert space is as usual the Hilbert space of the torus. Denote the holonomies of the gauge
field around the $A$- and $B$-cycles by $u$ and $v$, respectively. On the torus one can always conjugate
the gauge field into the Cartan subalgebra \cite{emss}, and so 
\be
[u_i,v_j]=g_s\d_{ij}
\ee
where $i$, $j$ run from 1 to $N$, and as usual $g_s=2\pi i/(k+N)$, thus incorporating the shift 
$k\rightarrow k+N$. On the solid torus, however, as explained in \cite{akmv}, large gauge transformations
that shift $v$ in the coroot lattice are not a symmetry, and so $u$ and $v$ are non-compact variables.
The partition function is then given by

\be
Z_{\sm{CS}}(S^3)=\bra0_v|e^{\mbox{\sm{Tr}}\, u^2/2g_s}|0_v\ket~,
\ee
where $|0_v\ket$ is the vacuum state, i.e. a solid torus with a trivial representation. Writing the
above in the $u$-basis then gives the matrix model expression for the partition function:
\be
\int\dd u_1\ldots\dd u_N\,e^{\sum_{i=1}^Nu_i^2/2g_s}\prod_{i<j}\sinh^2{u_i-u_j\over2}
\ee
up to normalization. 

It is useful to consider the mirror of this. It is given by resolving the conifold geometry
\be
xy=(e^u-1)(e^v-1)
\ee
on the open string B-model side by blowing up a ${\mathbb{P}}^1$ at $x=y=u=v=0$ \cite{candelas}.
One then has $N$ D1-branes wrapping the ${\mathbb{P}}^1$, whose action is given by the dimensional
reduction of the holomorphic Chern-Simons action to the world-volume of the D1's \cite{av,dv} which
describes the normal directions to the branes in the Calabi-Yau\footnote{For a careful treatment of 
such non-holomorphic embeddings, see \cite{dbdh}.}. As shown by Dijkgraaf and Vafa, this theory
reduces to a matrix model, whose solution gives the Riemann surface appearing in the deformed closed
string geometry \cite{dv}. This is well-known and we will not review it here.

More generally, if we have some gluing operator $U^{(p,q)}_{\l\m}\in SL(2,{\mathbb{Z}})$ which is built out the $S$ and $T$ modular matrices, one again obtains a matrix model on the $B$-model 
side. In Chern-Simons theory, the general expression for $U$ is known and it was given in \eq{U}.

\subsection{Brownian motion fermions and the B-model}

The Brownian motion problem is an intrinsically dissipative one. It is therefore natural to ask what is
the quantum mechanical system associated to it, that is, the quantum mechanical system whose amplitudes
are given by the Brownian motion probabilities. Since the probabilities are positive quantities and the
amplitudes are complex numbers, we do expect that some analytic continuation will be involved. In this
section we construct such a quantum mechanical system, and we find that it is given by a system of
fermions on a non-commutative space, which can be identified with B-model branes \cite{topvertex2,dst}.

Our aim is to set up a quantum mechanical problem such that
\be
p_t(\l,\m)=\bra\m|U(t)|\l\ket~.
\ee
We would like to find the states $|\l\ket$, and the operator $U(t)$ which gives the time evolution.
Keeping in mind the role of $\l$ and $\m$ as holonomies of the gauge field around the two cycles of
the torus and the canonical quantization of Chern-Simons reviewed in the previous section, it is natural 
to require
\be\label{commrel}
[\l_i,\m_j]=g_s\d_{ij}
\ee
but to still regard $\l_i$ and $\m_i$ as the coordinates of a system of $N$ particles, like in the 
original Brownian motion problem, since we are interested in describing motion of quantum mechanical
particles. Furthermore, recalling the non-intersecting condition, it is now natural to take the
particles to be fermions rather than bosons. That is, if we have a multi-particle state of the form
\be
\vf_1(\l_1)\vf(\l_2)\cdots\vf(\l_N)~,
\ee
we consider Slater wave-functions:
\be
\psi(x)={1\over\sqrt{N!}}\,\det(\vf_i(\l_j))_{1\leq i<j\leq N}~.
\ee
In what follows we will also use the short-hand notation
\be
|\l\ket={1\over\sqrt{N!}}\,\det(|\l_i\ket_j)~.
\ee
Of course, we should check that this is an orthonormal set of states of antisymmetric 
wave-functions. This is easy to check:
\be
\bra\l|\l'\ket=\det\d(\l_i-\l'_j)~.
\ee
We also have
\be
{1\over N!}\int_{{\mathbb R}^N}\dd^N\l\,|\l\ket\bra \l|\l'\ket=|\l'\ket
\ee
so really
\be
{1\over N!}\int_{{\mathbb R}^N}\dd^N\l\,|\l\ket\bra\l|=1~.
\ee
The factor of $N!$ comes from the natural measure induced by the integration over the Weyl chamber:
\be
\int_C\dd^N\l\,f(\l)={1\over N!}\int_{{\mathbb R}^N}\dd^N\l\,f(\l)
\ee
where $C$ is the set $\infty>\l_1>\l_2>\ldots>\l_N>-\infty$, and $f$ is a symmetric function.

Now we want to be able to represent \eq{commrel} in this fermionic Hilbert space. This is of
course just a Fourier transformation. For a single particle we have
\be
\vf_i(\m)={1\over\sqrt{2\pi g_s}}\int_{-\infty}^\infty\dd\l\,e^{\l\m/g_s}\vf(\l)~.
\ee
Obviously, we then have
\be\label{fourier}
\psi(\m)={1\over(2\pi g_s)^{N/2}}\int_{{\mathbb R}^N}\dd^N\l\,e^{{\sm{Tr}}\,\l\m/g_s}\psi(\l)
={1\over N!}{1\over(2\pi g_s)^{N/2}}\int_{{\mathbb R}^N}\dd^N\l\,\det(e^{\l\m/g_s})
\psi(\l)~.
\ee
We immediately compute
\be\label{lm}
\bra\l|\m\ket={1\over(2\pi g_s)^{N/2}}\det(e^{\l_i\m_j/g_s})~.
\ee
Fourier transforming back gives:
\be
\psi(\l)=(-2\pi/g_s)^{N/2}\int_{{\mathbb R}^N}\dd^N\m\,e^{-{\sm{Tr}}\l\m/g_s}\psi(\m)~.
\ee

We are now in a position to compute the operator $U(t)$. By definition, it will satisfy
\be
U(t+s)=U(t)U(s)
\ee
or in other words
\be
\bra\m|U(t+s)|\l\ket=\int_C\dd^N\nu\,\bra\m|U(t)|\nu\ket\bra\nu|U(s)|\l\ket~.
\ee
Computing $\bra\m|U(g_s)|\l\ket$, we find
\be
{\bra\l|U(g_s)|\m\ket\over\bra\l|\m\ket}=e^{-{\l^2+\m^2\over2g_s}}~.
\ee
Since this holds for any state, it also holds as an operator equation that
\be
U=e^{-\m^2/2g_s}e^{-\l^2/2g_s}~.
\ee
Of course, we could just as well define new states 
\be
\f(\l)=e^{-\l^2/2g_s}\vf(\l)~.
\ee
These transform as
\be
\f(\l)=(-2\pi/g_s)^{N/2}\int_{{\mathbb R}^N}\dd^N\m\,e^{-\l^2/2g_s}e^{-{\sm{Tr}}\l\m/g_s}\f(\m)~,
\ee
and their inner product gives the Brownian motion probability.

From \eq{lm} we see that the matrix elements $\bra\l|\m\ket$ again give the $SL(2,{\mathbb{Z}})$ 
$S$-matrix. This is a non-trivial fact and justifies the choice of fermionic wavefunctions. On the 
other hand, $U$ is a product of $T$-matrices $T_{\l\l}T_{\m\m}$. The latter is by construction so, as we 
demanded $p_t(\l,\m)=(TST)_{\l\m}$.

It is now natural to idenfity the wave-functions $\vf(\l)$ and $\vf(\m)$, which transform according to 
\eq{fourier}, as the B-model fermions of \cite{topvertex2}. In fact, in view of the discussion in the
previous subsection, $\l$ and $\m$ should really be regarded as as holomorphic coordinates rather than
real coordinates. This leads us to consider Brownian motion in the complexification of $U(N)$, that
is, to replace the $A_{N-1}$ Weyl chamber by its complexified version $gl_N$. That is, we are
considering $GL(N)$ rather than $U(N)$. In that case, $u$ and $v$ really describe coordinates on a
non-commutative ${\mathbb{P}}^1$, and the Brownian particles behave like B-model fermions when going
from one patch to the other. Indeed, they transform in a metaplectic representation of 
$SL(2,{\mathbb{Z}})$, and gluing one patch to the other with $SL(2,{\mathbb{Z}})$ transformations is done via Fourier transformation. This is true both for $\vf(\l)$, which transforms with $S$, and for
the redefined fermions $\vf(\l)$, which transform with $TST$. In a way, this closes the chain of 
dualities, since we started with Chern-Simons theory and we end up with the fermions describing branes
of the Kodaira-Spencer theory \cite{topvertex2,dst}. Notice that to achieve this we chose a {\it different}
representation of the Hilbert space from the one considered in the previous section, where we 
discussed the A-model.

The fact that $\l$ and $\m$ correspond to the holonomies of the gauge field around the cycles of the 
torus suggests that we should interpret the original Brownian motion description as some Brownian motion 
in field space. It would be interesting to see how this comes about in the path integral formulation.

\section{Matrix models for Chern-Simons and Brownian motion}\label{matrixmodel}

\subsection{One-matrix model and the Stieltjes-Wigert polynomials}

The matrix model for the partition function of Chern-Simons on $S^3$ with
gauge group $U(N)$ is \cite{marcos}:
\be\label{mm}
Z_{\sm{mm}}={1\over N!}\prod_{k=1}^N\int_{-\infty}^\infty
{\dd y_k\over2\pi}e^{-y_k^2/2g_s}\prod_{i<j}\left(2\sinh{y_i-y_j\over2}\right)^2~.
\ee
The matrix model of Chern-Simons in $S^3$ was solved in \cite{miguel} 
by the method of orthogonal polynomials. The key step was to realize that a coordinate 
transformation
\be\label{cotrafo}
y_i=\log x_i
\ee
brings it into a form for which the associated orthogonal polynomials are
explicitly known. With the transformation \eq{cotrafo}, the Gaussian potential
in \eq{mm} becomes a log-normal weight function $w(x)=e^{-\log^2x_i/g_s}$
\cite{szego}. On the other hand, the $\sinh$ Vandermonde repulsion term,
characteristic of unitary matrix models, becomes the usual Vandermonde
interaction $x_i-x_j$. The relation between the perturbation theory of the
Chern-Simons matrix model and topological invariants computed in a Hermitian
matrix model has been discussed in detail in \cite{marcos}. 

Notice that the quantum dimensions naturally appear in the expressions for the
Stieltjes-Wigert orthogonal polynomials \cite{szego}. Of course, this 
translates into the quantum dimensions that typically appear in expectation
values in Chern-Simons theory.

With these ingredients, the matrix model is easily solved and gives 
\cite{miguel}
\be\label{mm2}
Z_{\sm{mm}}=\left(g_s\over2\pi\right)^{N/2}e^{{1\over6}\,g_sN(N^2-1)}
\prod_{k=1}^{N-1}(1-q^{-k})^{N-k}
\ee
where as usual $q=e^{g_s}$. We should note here that the relation to the
partition function in canonical framing is\footnote{Note that, in contrast with the conventions 
in \cite{brownian}, we have included a factor of $e^{-\pi iN^2/4}$ in the 
definition of the Chern-Simons partition function in \eq{CS}. Also, in 
\cite{brownian} we dropped overall sign factors.} \cite{marcos}:
\bea\label{CS}
Z_{\sm{CS}}(S^3)&=&e^{-{1\over12}g_sN(N^2-1) -{i\pi\over4}\,N^2}\,
Z_{\sm{mm}}\nn
&=&{1\over(k+N)^{N/2}}\,\prod_{j=1}^{N-1}
\left(2\sin{\pi j\over k+N}\right)^{N-j}~,
\eea
where in the last line we used the standard relation between the topological
string coupling and the Chern-Simons coupling constant \eq{coupling}.

In what follows we will make more comments on
this matrix model --which is the same as the one for Brownian motion-- and the 
computation of more general expectation values.

Let us recall that the most characteristic property of the Chern-Simons matrix 
model is that it is a unitary matrix model with a Vandermonde interaction that 
is a $\sinh$ rather than the usual polynomial Vandermonde determinant. We will 
explore some interesting properties of this type of matrix model elsewhere 
\cite{sdhmt}. In a sense, this matrix model is much simpler than the ordinary 
ones because, in fact,  all the integrals involved in 
\eq{mm} are Gaussian after expanding the factors of $\sinh$, and in particular 
contain no polynomial terms, which would result in the usual logarithmic 
interactions. Thus they can actually be solved trivially without recourse to 
matrix models. The only non-trivial part is to keep track of the 
combinatorics, which can for example be done using the formulas in appendix 
\ref{extens}. Physically, the interesting part of this elementary exercise is
that it gives us a probabilistic interpretation of the partition function of Chern-Simons on
$S^3$. Indeed, in the appendix we prove that the Hopf link satisfies \eq{extensivity}. 
Now recalling the
identifications \eq{partfunction} and \eq{coupling}, we see immediately that,
if we set $x=y=\r$ and $s=t$, the left-hand side of \eq{extensivity} gives the 
value of the partition function of Chern-Simons on $S^3$ (with the appropriate 
framing as follows from \eq{partfunction}) as a function of the coupling, 
whereas the right-hand side gives the matrix model expression:
\be
e^{{1\over24}g_sN(N^2-1)}\,Z_{\sm{CS}}(S^3,g_s/2)={1\over N!}\,
e^{{1\over12}g_sN(N^2-1)}\int_{{\mathbb R}^N}{\dd^N\l\over(2\pi)^N}\,
e^{-|\l|^2/g_s}\prod_{i<j}\left(2\sinh{\l_i-\l_j\over2}\right)^2~.
\ee
Thus, if we set $x$ and $y$ to the vacuum, we see that the extensivity property
of the Brownian motion probability gives us a derivation of the matrix model
expression of the partition function of Chern-Simons. It is interesting to 
note that this partition function comes as a function of $g_s/2$ rather than
$g_s$. That is due to the fact that the left-hand side of \eq{extensivity} 
(with $s=t$) depends on $2t$ rather than $t$. This is reminiscent of 
a renormalization group equation. In fact, there is no need to set 
the external states to the vacuum. For generic external states $x$ and $y$, 
\eq{extensivity} can be rewritten in terms of expectation values of Wilson 
lines in the corresponding representations. Consider the more general formula
\eq{ppp} derived in the appendix (with $\b=0$):
\be\label{ppp0}
\int_C\dd^Nz\,e^{-{\a\over2}|z|^2}\,p_{t,N}(x,z)\,p_{s,N}(z,y)
=e^{-{\a\over2\D}({s\over t}|x|^2 +{t\over s}|y|^2)}\,p_{\D,N}(x,y)
\ee
where $\D=\a st+s+t$. Rewriting this in terms of (unnormalized) Wilson loops
(dropping phase factors)
\be
p_{t,N}(\l,\m)=e^{-{1\over2t}(|\l|^2+|\m|^2)}W_{\l\m}(t)
\ee
where we wrote the Hopf link as a function of $t$ rather than the level, we get
\be
e^{-{1\over2\Delta}({\a s(t-1)+s\over t}|\l|^2+{\a t(s-1)+t\over s}|\m|^2)}\,
W_{\l\m}(\Delta)=\int_C\dd^N\nu\,e^{-{\D\over2st}|\nu|^2}\,W_{\l\m}(t)
W_{\n\m}(s)~.
\ee
It would be interesting to find a field theory interpretation of this equation.

Let us add that, although using matrix models and the Stieltjes-Wigert 
polynomials to compute integrals of the type \eq{mm} might seem a little 
overdone, the power of these techniques should come to full use when computing 
more complicated observables, like those involving general torus links. We hope
to come back to this in the near future.

The probability of survival for $N$ Brownian movers was defined in 
\cite{fisher} as the probability that the $N$ movers survive after an amount
$t$ of time, or, in other words, that given some intial condition $\m$, after
time $t$ the trajectories of none of the movers crossed. Given that 
$p_{t,N}(\l,\m)$ is the probability of moving from $\m$ to $\l$ without 
intersecting, to get the probability that the movers have not intersected at
$t$ we have to integrate over $\l$:
\bea
p_{\sm{survival}}(\l)&=&\int_C\dd^N\l\,p_{t,N}(\l,\m)\nn
&=&{1\over N!}\int_{{\mathbb R}^N}\dd^N\l\,|p_{t,N}(\l,\m)|\nn
&=&{1\over N!}{1\over(2\pi t)^{N/2}} \int_{{\mathbb R}^N}\dd^N\l\,
e^{-{|\l|^2-|\m|^2\over2t}} |\det(e^{\l_i\m_j/t})|
\eea
where as usual the integral is over the Weyl chamber. In the second line we 
made use of the fact that $p_{t,N}$ is antisymmetric in $\l$. This can be 
approximated for large $t$ as follows \cite{grabiner}. We choose $t$ large and 
rescale $\l$ such that $|\l|/t$ is small but keeping $|\l|^2/t$ large, and take
the boundary condition so that $|\m|^2/t$ is small (and the particles are 
equally spaced). Keeping only linear terms, we get
\be\label{psurv}
p_{\sm{survival}}(\l)={1\over N!}{1\over(2\pi t)^{N/2}}\,t^{-N(N-1)/2}
\prod_{i<j}{\m_i-\m_j\over i-j}\int_{{\mathbb R}^N}\dd^N\l\,
e^{-{|\l|^2+|\m|^2\over2t}}\prod_{i<j}|\l_i-\l_j|~.
\ee
It is clear that the factor of $t^{-N(N-1)/2}$ came from the Vandermonde
interaction. It is amusing to note that this combines with the first term
to give a factor of $t^{-N^2/2}$. The above integral can be readily done using 
Selberg's integral \cite{mehtaboek} and is given in terms of $\G$-functions. 
After doing the integral, the total $t$-dependence is $t^{-N(N-1)/4}$ 
\cite{grabiner}. Expression \eq{psurv} was also found by Fisher \cite{fisher}.

Notice that from the above, for large $t$, we have
\be
p_{t,N}(\l,\m)\sim{1\over t^{N^2/2}}\,e^{-{|\l|^2+|\m|^2\over2t}}\Delta(\l)~.
\ee

This expansion can be used to compute more general observables:
\be
\bra f\ket={1\over N!}\int_{{\mathbb R}^N}\dd^N\l\,|p_{t,N}(\l,\m)|\,f(\l)
\ee
where we assumed that $f$ is symmetric in $\m$. One can again approximate
this for large $t$ as
\be
\bra f\ket={1\over N!}{1\over(2\pi t)^{N/2}}\,t^{-N(N-1)/2}
\prod_{i<j}{\m_i-\m_j\over i-j}\int_{{\mathbb R}^N}\dd^N\l\,
e^{-{|\l|^2+|\m|^2\over2t}}\prod_{i<j}|\l_i-\l_j|\,f(\l)~.
\ee
Again, for a large class of symmetric functions one can make use of Selberg's 
integral to evaluate this explicitly.

For antisymmetric functions, we write
\be
\hat f(\l)=\D(\l)\,f(\l)
\ee
where $f$ is symmetric. Then we simply get:
\be
\bra\hat f(\l)\ket={1\over N!}{1\over(2\pi t)^{N/2}}\, t^{-N(N-1)/2}
\prod_{i<j}{\m_i-\m_j\over i-j}\int_{{\mathbb R}^N}\dd^N\l\,
e^{-|\l|^2+|\m|^2\over2t}\D(\l)^2f(\l)
\ee
that is the Hermitian matrix model with $\b=2$. This is the usual case, and one can
use the results of for example \cite{zinn-justin} to compute the kernel. Its 
short-distance behavior is well-known:
\be
K(\l,\m)={\sin(N\pi(\l-\m)\r(\l))\over\pi(\l-\m)}
\ee
for $\l-\m\ll 1$. At this point one can make use of the usual matrix model
techniques to compute Brownian motion averages in the approximation of large $t$,
which corresponds to weak coupling on the field theory side.

\subsection{Brownian motion averages and the Hermitian two-matrix model}

We have seen how Brownian motion is related to one-matrix models. The 
probability $p_{t,N}(\l,\m)$ can however be interpreted more generally as an
integration kernel in the context of two-matrix models. For simplicity we will 
do this for the case of $U(N)$. Brownian motion two-matrix models have also
appeared in \cite{katori}.

We start with a simple example for the case $N=2$. Assume we are 
interested in the average of the final distance between the two particles, 
$\bra\m_1-\m_2\ket$ after time $t$, given some initial configuration 
$\l_1-\l_2$. As usual, we take $\l_1>\l_2$, $\m_1>\m_2$. We can compute this as
follows:
\be
\bra\m_1-\m_2\ket={1\over Z}\,\int_{\m_1>\m_2}\dd\m_1\dd\m_2\,p_t(\m,\l)
(\m_1-\m_2)
\ee
and in this case, explicitly,
\be
p_t(\m,\l)={1\over2\pi t}\,e^{-{|\l|^2+|\m|^2
\over2t}}(e^{\l_1\m_1+\l_2\m_2\over t}-e^{\l_1\m_2+\l_2\m_1\over t})~.
\ee
The normalization is given by the probability of survival
\be
Z=\int_{\m_1>\m_2}p(\m,\l)~.
\ee
So we get
\be
\bra\l_1-\l_2\ket={1\over Z}\int_{\m_1>\m_2}p(\m,\l)(\m_1-\m_2)=\l_1-\l_2~,
\ee
and $Z$ was computed asymptotically in \eq{psurv}. In what follows we will drop
such normalizations.

One can more generally consider averages of functions in both variables,
$f(\l,\m)$, corresponding to a process from $\l$ to $\m$ in time $t$, and one
may also be interested in integrating over both boundary conditions, initial
and final ones. Consider now for example:
\be
\bra {(\m_1-\m_2)(\l_1-\l_2)\over(1+\l_1^2)^2(1+\l_2^2)^2}\ket =2\pi^3
\ee
where we set $t=1$. From the above it is already clear that such integrals 
become more and more complicated as $N$ becomes large, and also one needs to 
make sure that they converge. In the above, the presence of the denominators
ensured convergence. Notice that in this case --but it holds more
generally-- after performing the $\m$-integrals all the Gaussian factors 
disappeared, and so in general one has
to consider rational functions rather than polynomials. 

In the following we will show how two-matrix models can be useful in this type
of computations, by mapping the density $p_t(\l,\m)$ to an integration kernel 
of a two-matrix model.

Thus we consider quantities of the following type:
\be\label{hatf}
\int_C\prod_{k=1}^N\dd\l_k\dd\m_k\,
\det(e^{-(\l_i-\m_j)^2/2t})_{1\leq i<j\leq N}\,\hat f(\l,\m)~,
\ee
and the integration region $C$ is as usual the fundamental Weyl chamber. The 
hat on $f$ is to indicate that it is an antisymmetric function under 
interchange of  any two $\l_i$ or $\m_j$. Notice that this is not necessary, as
antisymmetric functions do not vanish when integrated over the fundamental Weyl
chamber (in contradistinction to their vanishing when we integrate over the 
whole of ${\mathbb{R}}^N$). Nevertheless they are useful in that they can be
extended to integrals over ${\mathbb R}^N$ due to the identity:
\be
\int_C\prod_{i=1}^N\dd\l_i\,f(\l)={1\over N!}\int_{{\mathbb R}^N}\prod_{i=1}^N
\dd\l_i\,f(\l)~,
\ee
where $C$ is the fundamental Weyl chamber, for a symmetric total integrand $f$,
whereas for antisymmetric functions:
\be
\int_C\prod_{i=1}^N\dd\l_i\,\hat f(\l)={1\over N!}\sum_{w\in W}\e(w)\int_{w(C)}
\prod_{i=1}^N\dd\l_i\,\hat f(\l)
\ee
where $w(C)$ is the Weyl chamber reached by acting with an element of the Weyl 
group (in this case a permutation) on the fundamental Weyl chamber. Of course, 
we can always turn an antisymmetric function into a symmetric one by taking its
absolute value, and in that case one can extend the integral to 
${\mathbb R}^N$, as we saw in the case of the one-matrix model. For the time 
being we will limit ourselves to integrals over symmetric functions, that is, 
antisymmetric functions $\hat f$ in \eq{hatf}. For these ovserbables, the 
ordering of the particles is crucial. Without loss of generality we can write
\be
\hat f(\l,\m)=\Delta(\l)\Delta(\m)\,f(\l,\m)
\ee
where $\D(\l)$ is the usual Vandermonde determinant
\be
\Delta(\l)=\prod_{i<j}(\l_i-\l_j)~.
\ee
Thus we have
\bea\label{expval}
\bra\hat f\ket&=&N!\int_C\prod_{i=1}^N\dd\l_i\dd\m_i\,\Delta(\l)\Delta(\m)
\det(e^{-(\l_i-\m_j)^2/2t})\,f(\l,\m)\nn
&=&{1\over N!}\int_{{\mathbb R}^N}\prod_{i=1}^N\dd\l_i\dd\m_i\,\Delta(\l)
\Delta(\m)\det(e^{-(\l_i-\m_j)^2/2t})\,f(\l,\m)~.
\eea
Remember that $f(\l,\m)$ is a symmetric function. In the last expression, any
antisymmetric piece would not contribute. We will further assume that $\hat f$
is rational in $\l$, $\m$ and chosen such that the integral converges. 
We included the factor of $N!$ and dropped a factor of $1\over(2\pi t)^{N/2}$
(which we will nevertheless take into account later on) to simplify 
normalizations.

Before moving to the two-matrix model, it is convenient to rewrite the 
above in an alternative way. We note the following formula for arbitrary
symmetric function:
\be\label{expval2}
\int_{{\mathbb R}^N}\dd^Ny\,\Delta(y)\,\det(e^{x_iy_j})_{1\leq i<j\leq N}\,f(y)
=N!\int_{{\mathbb R}^N}\dd^Ny\,\Delta(y)\,e^{\sum_{i=1}^Nx_iy_i}\,f(y)~.
\ee
This is easily proven by writing out the definition of the determinant as a 
summation over the symmetric group, interchanging the order of summation and 
integration, and relabeling the $y$-coordinates. Similar manipulations are
illustrated in appendix \ref{extens}. This results in a summation
over $S_N$ that gives the factor of $N!$. Of course, the right-hand side is still
antisymmetric in the $x_i$'s, a property that will be crucial in what follows.

We can now rewrite expectation values of antisymmetric functions $\hat f$ as
\bea\label{brafket}
\bra\hat f\ket&=&\int_{{\mathbb R}^N}\prod_{i=1}^N\dd\l_i\dd\m_i\,\Delta(\l)
\Delta(\m) e^{-\sum_{i=1}^N(\l_i-\m_i)^2/2t}\,f(\l,\m)~.
\eea
Again, any antisymmetric piece in $f$ drops out after symmetrization.

It is not hard to recognize in \eq{expval} a two-matrix model. The form
\eq{expval} is the one used in \cite{itzzub} which results from the
Itzykson-Zuber integral, the simpler one \eq{brafket} is the one advocated in 
\cite{mehta}.  It is a two-matrix model interacting via
a potential $\sum_{i=1}^Nx_iy_i$, as we will now see.

Consider then the Hermitian two-matrix ensemble \cite{itzzub}:
\be\label{2mm}
Z=\int[dL][dM]\,e^{-V(L)-V(M)+\b\sm{Tr}(LM)}~.
\ee
Averages in this ensemble are integrals of the form
\be
\bra f(L,M)\ket=\int[dL][dM]\,e^{-V(L)-V(M)+\b\sm{Tr}(LM)}f(L,M)~.
\ee
Typically, $f(L,M)$ is a rational function of the moments $\Tr L^m$, $\Tr L^n$.

We will clearly be interested in quadratic potentials\footnote{Notice, however,
that in this special case $Z$ is not well defined: it diverges like 
$\sim\Lambda^{N^2}$ where $\Lambda$ is a cutoff on the $L,M$-integrals, or,
alternately, as $\sim1/(\b-1/t)^{N/2}$ \cite{kazmar} as $\b\rightarrow1/t$.
This is however not the Brownian motion normalization.} $V(M)=M^2/2t$ and 
$\b=1/t$, in which case the above takes the form:
\be
\bra f\ket=\int[dL][dM]\,e^{-\sm{Tr}(M-L)^2/2t}f(L,M)~,
\ee
for some function $f(L,M)$. For now, however, we will leave the parameters
generic. The measure $[dL]$ has been worked out in \cite{itzzub,mehta} and it 
contains two parts: an integral over the diagonal entries of $L$, and an
integral over the angular part. Indeed, we can diagonalize $L$ by a unitary
matrix $U$ as
\be
L=U\,\Lambda U^\dagger~.
\ee
where $\Lambda={\mbox{diag}}(\l_1,\ldots,\l_N)$. This, however, cannot be done 
simultaneously for $L$ and $M$. Thus there 
remains an integral over unitary matrices that we have to perform explicitly.
We have \cite{mehta}
\be\
\dd L=\dd\Lambda\dd U\,\Delta^2(\l)~,~~~~\dd\Lambda=\prod_{i=1}^N\dd\l_i~.
\ee
References \cite{itzzub,mehta} then prove that the above reduces to
\be
\bra f\ket={(2\pi)^{N(N-1)}\over(\prod_1^Np!)^2}\int\prod_{i=1}^N\dd\l_i
\dd\m_i\,\Delta(\l)\Delta(\m)e^{-V(\l)-V(\m)}I(\l,\m)f(\l,\m)~,
\ee
and we are assuming that $f(L,M)$ is a function of the moments of $L$, $M$ and
no mixing terms. $I(\l,\m)$ is the Itzykson-Zuber integral, given by
\be\label{IZint}
I(\l,\m;t)=t^{N(N-1)/2}\prod_1^Np!{\det(e^{\l_i\m_j/t})\over\Delta(\l)
\Delta(\m)}
\ee
and it is the result of the integration over the angular variables. We see that
up to normalization this is precisely the integral\footnote{In this formula $M$
denotes the diagonal entries of $M$ and not $M$ itself.} \eq{expval}:
\be
\int[dU]\,e^{-{1\over2t}\sm{Tr}(\L-UMU^\dagger)^2}=t^{N(N-1)/2} \prod_1^{N-1}
p!\,{\det(e^{-(\l_i-\m_j)^2/2t)}\over\Delta(\l_i)\Delta(\m_i)}~.
\ee
Putting everything together, this gives the Brownian motion ensemble 
\eq{expval}.

One can use uniqueness of the solution of the heat equation with respect
to the unitary invariant Laplacian operator on Hermitian matrices and on the
Cartan subalgebra, respectively, to compute the coefficient appearing in the
transformation from $\dd M$ to an integral over the diagonal. What is of 
interest for us is the fact that the
solution kernel of the former equation is
\be
f(L,M;t)={1\over(2\pi t)^{N^2/2}}\,e^{-{\sm{Tr}}(L-M)^2/2t}~,
\ee
and this explains the $t$-dependence in \eq{IZint}. Let us now turn to the
heat equation on the Cartan subalgebra. This is of course the heat equation
we considered before:
\be
{\pa\over\pa t}f=\half \Delta f~,
\ee
and we look for antisymmetric solutions of this equation, 
$f(\l_1,\ldots,\l_N)$ with 
boundary condition $f_0(\l)$ at $t=0$. These are given by \cite{itzzub}
\be
f(\l)=\prod_{i=1}^N\int\dd\m_i\,K(\l,\m;t)\,f_0(\m)
\ee
and
\be
K(\l,\m;t)={1\over(2\pi t)^{N/2}}{1\over N!}\,\det(e^{-(\l_i-\m_j)^2/2t})
\ee
is precisely the Brownian motion density $p_{t,N}(\l,\m)$. Of course, any 
{\it symmetric} solution of the heat equation can be obtained by rescaling the
corresponding symmetric solution with the Vandermonde determinant.

This is of course analogous to the case of Brownian motion in the affine 
Weyl chamber of $U(N)$/two-dimensional Yang-Mills \cite{sdh}, where the 
partition function is the heat kernel of solutions of the heat equation 
for symmetric functions.

Having found a correspondence between Brownian motion averages and one- and 
two-matrix models, we can now use the techniques available for these models, in 
particular large-$N$ techniques. In particular, we have the following correlation
function:
\be
\r_N(\l_1,\ldots,\l_N;\m)={1\over N!}{1\over Z}\,\Delta(\l)\,
\det(e^{-(\l_i-\m_j)^2/2t})
\ee
in terms of which we can define our Brownian motion averages for antisymmetric
functions, as before. The normalization factor is
\be
Z=\int\dd^N\l\,\D(\l)\det(e^{-(\l_i-\m_j)^2/2t})~.
\ee
This is the analog of the density of eigenvalues in the presence of a coupling to
an external source $\m$. It is related to the two-point kernel as follows:
\be\label{detK}
\rho_N(\l_1,\ldots,\l_N)={1\over N!}\det\underline{K}(\l_i,\l_j)~.
\ee
We supressed explicit $\m$-dependence because here we are considering it as a 
source, but these expressions depend on $\m$. An explicit expression for 
$\underline{K}$ has been computed in \cite{zinn-justin},
and the large-$N$ asymptotics and short distance behavior have been analyzed in
\cite{zinn-justin,johansson}. It is well-known that $K$ can be expressed in terms of
orthogonal polynomials. The generalization of this to the present case is in 
\cite{zinn-justin}. One can also integrate out some of the coordinates $\l_i$ to get a correlation function depending on a smaller number $n$, $n<N$, of 
eigenvalues. In that case the determinantal identity \eq{detK} continues to hold
\cite{zinn-justin}.

Notice that this kernel is not the kernel computed
before; symmetric solutions of the heat equation can be obtained from it, and it 
satisfies \eq{extens1d}. In fact, it is related to the Brownian motion kernel as
follows:
\be
\det\underline{K}(\l_i,\l_j;\m)={1\over\int\dd^N\l\,\D(\l)K(\l,\m)}\,\D(\l)
K(\l,\m)~.
\ee

Two-matrix models are solved in \cite{itzzub,mehta}. \cite{bereyn} has found
exact formulas for mixed correlation functions of the type $\Tr L^nM^m$
using biorthogonal polynomials.

Notice that the same two-matrix model was conjectured to play a role in the computation of correlators of branes in Calabi-Yau crystals \cite{dst}. Since in the previous section we already saw that Brownian motion particles can be interpreted as B-model fermions, it would be interesting to see if the above formulas can be used in the context of \cite{dst}, and whether they have a Chern-Simons interpretation.

\section{Discussion and outlook}

In this paper we have shown that for a number of observables the mathematical structures of Chern-Simons theory and Brownian motion for 
non-intersecting movers are almost identical, for the $U(N)$, $SO(N)$ and $Sp(N)$ groups and for the manifolds $S^3$ and $S^3/{\mathbb{Z}}_p$. The reformulation of the 
classical Brownian problem as counting of quantum mechanical paths in the Weyl chamber of the gauge group points to a deeper connection between 
Chern-Simons and Brownian motion, namely at the level of the path integral where we count paths on field space. Indeed, from the canonical 
quantization approach it became clear that the coordinates of the particles describe the holonomies of the gauge field around the cycles of the torus along which one does the gluing.

An intriguing connection with the B-topological string was found at the level of fermions. The natural fermionic formulation that we find for the Brownian particles is identical to the one in \cite{topvertex2}, thus identifying our particles with the branes of the B-model. In the case of $S^3$ that we analyzed, we found that these fermions move on a non-commutative space with two patches. For this we needed to complexify the Weyl chamber, thus effectively replacing $U(N)$ by $GL(N)$. Brownian propagation then translates into a coordinate transformation from one patch to the other which is a Fourier transformation. 

The case of lens spaces corresponds to motion in the affine Lie algebra, and is related to 2d Yang-Mills theory. \cite{sdh} related Chern-Simons on lens spaces to 2d Yang-Mills on the cylinder at a value of the coupling that is $2\pi i$ times a rational number. However, the results in \cite{aosv} suggest to consider also the analytically continued Chern-Simons. Since integer values of the coupling are required in Chern-Simons in order to preserve invariance under large gauge transformations, the deformed theory is defined by the quantum deformed 2d Yang-Mills, which for the case of the sphere is the same as usual 2d Yang-Mills on the cylinder, as we showed. Thus, the modular transformation properties worked out in \cite{sdh}
should  correspond to those of the A-model on the corresponding bundle over the two-sphere.
It would be interesting to investigate this further. In fact, Brownian motion would seem to be more naturally associated with the quantum-deformed 2d Yang-Mills than with ordinary Yang-Mills, due to the exponential nature of the probabilities. This quantum deformed 2d YM \cite{aosv} is an interesting theory in its own right. The partition function was evaluated in \cite{sdh,aosv} for the sphere and the torus, and by using the manipulations in appendix \ref{extens} it should be readily extendable to the higher genus case. In particular, one can obtain matrix model expressions, which are useful to work out the large $N$ limit. We will come back to this in the future \cite{sdhnew}.

It is likely that the generalization of the general modular matrix $U^{(p,q)}_{\l\m}$ is simply 
given by the same expression \eq{U}, where one analitically continues in the coupling and sums over the whole 
coroot lattice. It would be interesting to check this. In that case one can use the relation to
the $\Theta$-functions to study its modular transformation properties, as in \cite{sdh}, and one could
study the reformulation of Chern-Simons theory in terms of 2d qYM for more general manifolds.

One important issue is to extend the Brownian motion description of Chern-Simons to more general manifolds and knot invariants. Particularly interesting is the case of knots consisting of three links or more, since so far we encountered only up to two-linked knots. This is of course due to the fact that we only have two boundary conditions to our Brownian motion problem. This suggests that for extra links -- which will carry additional representations -- one will have to impose more boundary conditions on the problem. One could for example consider the conditional probability of starting with a configuration $\l$, ending up with a configuration $\m$, {\it given} that the intermediate situation at some time $t$ was an intermediate state $\n$. It would be interesting to check whether this associates an invariant with representations $\l,\m,\n$ to a three-link knot. It would be very interesting to work along these lines further, as this might give an independent and general way of computing knot invariants. It might be useful here to really view the Brownian motion picture as a braid diagram, and consider the associated algebra.

An interesting open problem is to develop matrix model technology further so one is able to compute for example general torus links, and in particular their large $N$ expansions. The same is true for three-manifolds $M$ where $M$ is a Seifert homology sphere, so one can work out the corresponding geometric transitions on $T^*M$. As far as Brownian motion is concerned, we saw that in order to replace $S^3$ by $S^3/{\mathbb{Z}}_p$ one had to replace the finite fundamental Weyl chamber by the affine Weyl chamber, which amounts to modding out by translations in the coroot lattice. It would be interesting to generalize this procedure to other manifolds, and in particular see if there is a general way of modifying the manifold by modifying the shape of the region where the particles are allowed to walk.

Recently, the melting crystal picture was applied to Chern-Simons on $S^3$ directly \cite{okuda}. It should be possible to see directly what melting crystals and Brownian motion have to do with each other.

Finally, it would be interesting to understand the role of the heat equation within Chern-Simons theory (perhaps in terms of stochastic quantization \cite{spenta}, as a Fokker-Planck equation) and in the topological string itself.

\section*{Acknowledgements}

\addcontentsline{toc}{section}{Acknowledgements}

We thank Robbert Dijkgraaf, Axel Kleinschmidt, Sergei Nechaev, Ingo Runkel, Gerard 't Hooft, Spenta Wadia, and especially Marcos Mari\~no for many discussions and useful comments. We also thank the theory division of CERN and ENS, Paris, for hospitality during the completion of this work.

\appendix

\section{Extensivity properties of the kernel}\label{extens}

In this appendix we give the proof of the extensivity formula \eq{extensivity}:
\be\label{extensivity2}
p_{t+s,N}(x,y)=\int_C\dd^Nz\,p_{t,N}(x,z)\,p_{s,N}(z,y)
\ee
and work out several other identities of the same type.
 We first write $p_{t,N}(x,y)$ as:
\be
p_{t,N}(x,y)= {1\over(2\pi t)^{N/2}}\,\sum_{\s\in S_N}\e(\s)\, e^{-\sum_{i=1}^N(x_i^2+y_i^2-2x_iy_{\s(i)})/2t}
\ee
so that the right-hand side of \eq{extensivity2} is
\be
{1\over N!}{1\over(4\pi^2st)^{N/2}}\sum_{\s,\s'}\e(\s\s')\, \int_{{\mathbb R}^N} \dd^Nz\, e^{-\sum_{i=1}^N[(z_{\s(i)}-x_i)^2/2s+(z_{\s'(i)}-y_i)^2/2t]}~.
\ee
We now relabel the $z_i$'s and perform a transformation $\s\rightarrow \s'\s$. The $\s'$ sum then results in a factor of $N!$. Bringing the remaining terms together and performing the Gaussian integrals, we are left with:
\be
{1\over(2\pi(s+t))^{N/2}}\sum_{\s\in S_N}\e(\s)e^{-\sum_{i=1}^N(x_i-y_{\s(i)})^2/2(s+t)}=p_{s+t,N}(x,y)~,
\ee
which is what we wanted to prove.

In fact we can easily extend this computation to the following:
\bea\label{ppp}
&&\int_C\dd^Nz\,e^{-{\a\over2}|z|^2+\b\sum_{i=1}^Nz_i}\,p_{t,N}(x,z)\,
p_{s,N}(z,y)\nn
&&=e^{-{1\over2\D}\left[\a({s\over t}|x|^2 +{t\over s}|y|^2)
-2\b\sum_{i=1}^N(sx_i+tz_i)-N\b^2ts\right]}\,p_{\D,N}(x,y)
\eea
where $\D=\a st+s+t$. We can also replace $\b\sum_{i=1}^Nx_i$ by 
$\sum_{i=1}^N\b_ix_i$ in the integrand, and then the final expression contains
an additional symmetrization with respect to $\b_i$.

Notice that we can construct $q_{t,r}(\l,\m)$ -- or, equivalently, the partition function of Chern-Simons on lens spaces -- from two $S^3$ partition functions. Indeed, up to normalizations, we have
\be
q_{t',r}(\l,\m)=\left({\a\over2\pi}\right)^{r/2}\sum_{n\in Q^\vee/pQ^\vee} \int_{{\mathbb{R}}^N}\dd^rz\,e^{-{\a\over2}z_i^2+n_iz_i/t}p_{t,r}(\l,z)p_{t,r}(z,\m)
\ee
where the coupling is $t'=\a^2t$. The left-hand side is the operator $ST^qS$. One can also show extensivity of $q_{t,r}(\l,\m)$ itself. Analogous expressions can be derived for q-deformed 2dYM.

\section{Path integral formulation of the path counting}\label{pathi}

Computing non-intersecting Brownian motion probabilities involves picking an infinite set of non-intersecting paths out of the infinite set of all free Brownian motion paths. It is therefore natural to look for a path integral formulation, where particles are traveling from an initial state to a final state along Brownian paths. It is well-known that free Brownian paths correspond to quantum mechanical paths, in other words, the free Brownian motion probability is the quantum mechanical path integral of a free particle. Thus, all we have to do is take into account the fact that the particles are non-intersecting. By elementary manipulations, we get:
\be
p_{t,N}(\l,\m)=\det\int_{\l_i}^{\m_j}{\cal D}x_1(t)\ldots{\cal D}x_N(t)e^{iS[x(t)]}
\ee
where the determinant is taken with respect to the boundary conditions $\l_i$ and $\m_j$, and $S[x(t)]$ is the action for a free particle. It is interesting to discretize this as follows:
\bea
&&p_{t,N}(\l,\m)=\nn
&&=\lim_{\e\rightarrow0}\left(m\over2\pi i\e\right)^{MN/2}\int_C\dd x^2_1\ldots\dd x^2_N\ldots
\int_C\dd x_1^{M-1}\ldots\dd x_N^{M-1}\prod_{i=1}^{M-1}\det(e^{{im\over2\e}(x_a^i-x_b^{i+1})^2})_{ab}~,
\eea
where we simultaneously take the limit $\e\rightarrow0$ and $M\rightarrow0$. The coordinates $x^a_i$ label the discretized coordinates, where $i=1,\ldots,N$ and $a=1,\ldots,M$. Notice that we were able to restrict all integration regions to the Weyl chamber, as expected. Thus, this can be seen as a path integral for motion inside the Weyl chamber.

It is likely that the above expression presents an additional hidden supersymmetry when expressing the determinant as a fermionic integral, in the limit $M\rightarrow\infty$, as in \cite{parisisourlas}. It would be interesting to work this out further.

\section{Some determinantal identities}\label{vdmonde}

In this appendix we collect the steps leading to \eq{prob00}. We use the
following identities:
\bea\label{Vdm}
\det(e^{\l_i(j-1)})&=&\prod_{1\leq i<j\leq N}(e^{\l_j}-e^{\l_i})\nn
\prod_{1\leq i<j\leq N}q^{\l_j}&=&q^{\sum_{i=2}^N(i-1)\l_i}~.
\eea
The first is the standard Vandermonde identity, and the latter is also easily 
Checked.

We start from the last equality in \eq{ptN} with $\l_i=\m_i=(c-i)a$ and extract
the Gaussian factors:
\be
\det(e^{-{(\l_i-\m_j)^2\over2t}})_{1\leq i<j\leq N}
={1\over(2\pi t)^{N/2}}\,e^{-\sum_{i=1}^N(i-1)^2a^2/t}
\det(e^{(j-1)(i-1)a^2/t})~,
\ee
where we used the freedom to shift the initial and final positions to recast
the determinant in a form in which we can apply the standard Vandermonde
determinant formula \eq{Vdm}. Then using both of \eq{Vdm}, we see that the
prefactors cancel and we finally get
\be\label{det}
\det(e^{-(j-i)^2a^2/2t})=\prod_{k=1}^N(1-q^k)^{N-k}
\ee
with $q=e^{-a^2/t}$.

For the computation of the expectation value of a Wilson we need in addition
the identity
\be\label{det2}
\prod_{i<j}(e^{x_j}-e^{x_i})=\prod_ke^{{N-1\over2}\,x_k}
\prod_{i<j}2\sinh{x_j-x_i\over2}~.
\ee
This is easyly checked for low values of $N$, and proved by 
induction.

\section{Representations and partitions}\label{partitions}

We first introduce some Lie algebra conventions. In this appendix we deal 
with the case of $U(N)$ and $SU(N)$. We start with $U(N)$. We use standard 
notations (see for example \cite{diF}), where $\o_i$, 
$i=1,\ldots,N$, denotes the set of fundamental weights, and $\e_i$ are unit
vectors in ${\mathbb R}^N$. They are related by:
\be
\o_i=\sum_{j=1}^i\e_j -{i\over N}\sum_{j=1}^N\e_j.
\ee
The simple roots are given by $\a_i=\e_i-\e_{i+1}$ for $i=1,\ldots,N-1$. From 
here it follows that 
$|\a|^2=2$ and so $\a=\a^\vee$. Now we can 
expand a highest weight of an irrep
\be\label{loi}
\l=\sum_{i=1}^N\l_i\o_i=\sum_{i=1}^N(\ell_i-\k)\e_i
\ee
where
\be
\k={1\over N}\sum_{j=1}^Nj\l_j
\ee
is the number of boxes in the Young tablaux divided by $N$, and
\be
\ell_i=\l_i+\l_{i+1}+\ldots\l_N~.
\ee
Using this definition, it is not hard to check that \eq{loi} holds.

A Weyl chamber is a connected set in 
${\mathbb R}^N$ left out after we delete all hyperplanes orthogonal to the 
roots. It is defined by
\be
C_w=\{\l|(w\l,\a_i)\geq0,i=1,\ldots,r\},\,\,w\in W~.
\ee
The fundamental Weyl chamber corresponds to the identity element and is denoted $C_0$. Obviously, 
$\l_i\geq0$ for all $i=1,\ldots,N$ if $\l$ is in the fundamental Weyl chamber. This implies that
the $\ell$'s satisfy $\ell_i\geq\ell_{i+1}$.  Defining coordinates
\be
h_i=\ell_i +{N+1\over2}-i~,
\ee
we see that the fundamental Weyl chamber is the domain $h_1>h_2>\ldots h_N$.
These are the variables that are usually used to label representations, and
which we have labeled by $\l$ and $\m$ (and we will continue to do
so). They can be written as
\be
h=\ell+\r~,
\ee
where $\r$ is the Weyl vector
\be
\r=\sum_{i=1}^N\o_i=\sum_{i=1}^N\left({N+1\over2}-i\right)\e_i
\ee
and has norm
\be
|\r|^2={1\over12}\,N(N^2-1)~.
\ee

The fundamental weights satisfy:
\be
(\o_i,\o_j)=F_{ij}=(C^{-1})_{ij}{|\a|^2\over2}=(C^{-1})_{ij}
\ee
where in the last line we specialized to $U(N)$. $C$ is the Cartan matrix. The
inner product for weights is
\be
(\l,\m)=\sum_{ij}\l_i\m_jF_{ij}~.
\ee

For $U(N)$, the Casimir becomes especially simple in the new basis:
\bea
C_2(\l)&=&(\l,\l+2\r)\nn
&=&|\ell+\r|^2-|\r|^2=|h|^2-|\r|^2=\sum_{i=1}^Nh_i^2-{1\over12}\,N(N^2-1)~.
\eea

The dimension of a representation is given by
\be
{\mbox{dim}}\,R=\prod_{1\leq i<j\leq N}{h_i-h_j\over j-i}
\ee

%For $U(N)$ we also have $g=N$, $\dim g=N^2-1$, $r=N$, $|\D_+|=\half N(N-1)$
%and $|P/Q^\vee|=N+1$.

\section{A note on framing}\label{framing}

In previous sections we related Brownian motion and Chern-Simons quantities. 
Framing in Chern-Simons theory was already explained in section \ref{CStheory}.
Both in the case of the partition function and for the unknot we saw that
the expressions resulting from Brownian motion were not quite in canonical 
framing, but in matrix model framing. That is, every state came multiplied with
a Boltzmann factor. We will now explain how such factors can be modified by 
changing the boundary conditions of the Brownian motion. We will consider 
translations and dilatations of the boundary conditions. Besides the possible 
interpretation as framing factors in Chern-Simons, these properties of the 
probability densities are interesting in their own right.

We start with an example that will however cover all of the physical effects
we want to analyze. We take initial and final states that are equally spaced,
but are not identical:
\bea
\l_i&=&(c-i)a\nn
\m_i&=&(c'-i)a'~.
\eea
Let us further introduce the notation
\bea\label{newbc}
\a&=&ca-c'a'\nn
\b&=&a-a'~.
\eea
We are thus set to compute the determinant \eq{ptN} for the above values of 
$\l,\m$. This is not hard to do and the procedure is
similar to the way we found the probability of reunion. Basically, we expand
the exponential in such a way that we can use formula \eq{det}. The result is
\be\label{dila}
\det(e^{-(\l_i-\m_j)^2/2t})=e^{[-{1\over6}\b^2N(N+1)(2N+1)-N\a(\a+\b(N+1))]
{1\over2t}}\prod_{k=1}^{N-1}(1-e^{-kaa'/t})^{N-k}~.
\ee

We see that $\a$ and $\b$ can be chosen such that the exponential equals a 
framing factor ${\pi ik(N^2-1)\over12(k+N)}n$ for some integer $n$. So in 
principle we can reach any framing we want by translating the initial or the
final positions of all the particles by a constant factor, or by rescaling
them. Let us however stress that these are not {\it symmetries}: the 
probability densities change by exponential factors. However, once we have 
computed the probability for a specific configuration, we can compute that of
any other configuration connected to it by translations or dilatations with
the above transformation rules.

Not all of the above transformations have a natural interpretation in terms of
framing though; the translations do not. To see this, it is best to look not
at the ``probability of reunion''\footnote{We will continue to use this name,
even though with the boundary conditions \eq{newbc} the particles do not
go back to their initial positions, now even up to an overall translation.}, 
but more generally to the transformation of $p_{t,N}(\l,\m)$ under 
dilatations and translations. Let us first consider the latter. We already know
that the probability density is invariant under translations by equal factors, 
$\l\rightarrow\l+c$, $\m\rightarrow+c$. We now consider the following:
\be
p_{t,N}(\l+ca,\m+c'a)=e^{-\sum_{i=1}^N(\l_i-\m_i)(c-c'){a\over t}
-N(c-c')^2{a^2\over2t}}\, p_{t,N}(\l,\m)~.
\ee
Clearly, the first factor does not have much chance of being interpreted as a
framing factor except in an $SU(N)$ theory where it vanishes. For $U(N)$, the framing factors go with the second 
Casimir of the representation, not with the first (see section \ref{CStheory}).
Of course, this factor vanishes when $|\l|=|\m|$, and in particular for the
Weyl vector, which satisfies $|\r|=\sum_{i=1}^N\r_i=0$ (see appendix 
\ref{partitions}). But there is no reason to restrict oneself to this type of
boundary conditions, so this does not generally reproduce the effect of framing.

The effect of dilatations on the external states however is much more similar
to framing. This corresponds in \eq{newbc} to $\a=\b$. Let us now again
consider the most general case of arbitrary initial and final states. To this
end it is easiest to work directly with \eq{ptN}. If we rescale
\bea
\l&\rightarrow&\l a\nn
\m&\rightarrow&\m a'~,
\eea
we get
\be
p_{t,N}(\l a,\m a')={1\over(2\pi t)^{N/2}}\,e^{-{|\l|^2\over2t}a^2
-{|\m|^2\over2t}a'^2}\det(e^{\l_i\m_jaa'/t})~.
\ee

Of course, this reduces to \eq{dila} for $\a=\b$ and the probability of 
reunion. However, we do not want to change the last interaction term, which
corresponds to the modular matrix $S$ and does not change under framing. At
this point we can either generalize the relation between $t$ and the 
Chern-Simons coupling to include the extra factors of $a,a'$, or we can
consider rescalings of time as well:
\be
p_{t',N}(\l',\m')={1\over(2\pi aa't)^{N/2}}\,e^{-{|\l|^2\over2t}a^2
-{|\m|^2\over2t}a'^2} \det(e^{\l_i\m_j/t})~.
\ee
This now corresponds, up to an overall constant, to a matrix element
\be
\bra\m|T^{n}ST^{m}|\l\ket
\ee
which is what we wanted. For this to be a framing, $a^2=n$ and $a'^2=m$ are
integers. With $a$ and $a'$ real, notice that this can only {\it augment} the
framing and not diminish it.

\section{Partition function of $SO$ and $Sp$ Chern-Simons}\label{SOSpCS}

\subsection{$SO(2N+1)$ and $Sp(2N)$}

Let us now give $f(k)$. For 
$SO(2N+1)$:
\be\label{fso1}
{f(k)= \cases {
                       1 & $k=j-\half$ ($j=1,\ldots,N$)     \cr
                       N-k/2 &  $k$ even \cr
                       N-k/2-\half & $k$ odd~.}}
\ee
For $Sp(2N)$:
\be\label{fsp}
{f(k)= \cases {
                       N-k/2-\half & $k$ odd $\leq N$     \cr
                       N-k/2 &  $k$ even $<N$             \cr
                       N-k/2+\half & $k$ odd $>N$          \cr
                       N-k/2+1     & $k$ even $>N$~.}}
\ee

\subsection{$SO(2N)$}

For $SO(2N)$:

Again, using Weyl's denominator formula this can be brought to the form
\bea
Z_{\sm{CS}}(S^3)&=&S_{00}=(k+g)^{-N/2}\prod_{i<j}\sinh{j-i\over2t}
\sinh{2N-j-i\over2t}\nn
&=&(k+2N-2)^{-N/2}\prod_{k=1}^{N-1}(\sinh k/2t)^{f(k)}
\prod_{k=N}^{2N-1}(\sinh k/2t)^{f(k)-1}\nn
&=&(k+2N-2)^{-N/2}\prod_{k=1}^{2N-1}(\sinh k/2t)^{f(k)}/
\prod_{k=N}^{2N-1}\sinh k/2t
\eea
and $f(k)=N-1/2k$ if $k$ is even, and $N-1/2k+1/2$ if it is odd, in other words,
\be
f(k)=[N-{k+1\over2}]~.
\ee

\end{document}